\documentclass[11pt]{article}

\usepackage[utf8]{inputenc}
\usepackage[T1]{fontenc}
\usepackage{lmodern}
\usepackage{csquotes}
\usepackage[UKenglish]{babel}
\usepackage[a4paper, margin=1in]{geometry}
\usepackage{setspace}
\usepackage[protrusion=true,expansion=true]{microtype}
\usepackage{newpxtext,newpxmath}

\usepackage{amsmath}
\usepackage{bm}

\usepackage{graphicx}

\usepackage{booktabs}
\usepackage{threeparttable}
\usepackage{tabularx}
\usepackage{multirow}
\usepackage{array}
\usepackage{dcolumn}
\usepackage{caption}
\usepackage{subcaption}
\usepackage{float}
\usepackage{rotating}
\usepackage{pdflscape}
\usepackage[hidelinks]{hyperref}
\usepackage{appendix}
\usepackage{xcolor}
\usepackage{csvsimple}
\usepackage{siunitx}

\graphicspath{figures/}

\usepackage[
    backend=biber,
    style=apa,
    sorting=nyt,
    maxcitenames=2,
    mincitenames=1,
    url = false,
    doi = true,
    dashed=true,
    date=year]{biblatex}
\title{\textbf{From Exposure to Adoption:\\
Generative AI in European Workplaces}}
\date{\today}

\author{Golo Henseke\footnote{University College London, 20 Bedford Way,
London WC1H 0AL, UK, email: \href{mailto:g.henseke@ucl.ac.uk}{g.henseke@ucl.ac.uk}}}
\begin{document}
\maketitle

\begin{abstract}
\noindent
This study examines who adopts generative AI and whether early adoption has begun to reshape the task content of jobs across 35 European countries. Adoption ranges from under 3\% to 25\%. Occupational exposure strongly predicts uptake, but AI does not diffuse passively along exposure lines. At the worker level, skills, abstract task content, and employee organisational influence steepen the exposure--adoption gradient; at the country level, so do digitalisation and workplace training. A gender gap persists, concentrated in the most exposed occupations. A shift-share design finds no detectable effect of adoption on worker-reported task restructuring, consistent with an initial integration phase.
\end{abstract}


\medskip
\bigskip

\section{Introduction}
\label{sec:intro}

The launch of ChatGPT in November 2022 triggered rapid diffusion of a novel general-purpose technology, potentially outpacing the adoption of personal desktop computers in the mid-1980s \parencite{Bick2024TheAI}. By early 2024, ChatGPT alone attracted two billion monthly visits and 500 million unique users globally \parencite{Liu2026WhoAI}. 26\% of U.S. workers reported using generative AI on their job in late 2024, rising to 43\% by early 2026 \parencite{Bick2024TheAI,Bick2026MindU.S.}. In the UK, the share of workers using AI software at work surged from 15\% to 24\% between Q3 2023 and Q2 2024 \parencite{Henseke2025WhatAdoption}. By 2025, across Europe, 3 in 10 workers may have used AI-powered tools \parencite{GonzalezVazquez2025DigitalEurope}. These aggregate statistics, however, mask enormous heterogeneity in who adopts, where, and under what conditions.

Three dimensions of this heterogeneity are particularly consequential. First, adoption is unequal across occupations: workers in roles with greater task-level exposure to AI capabilities are more likely to adopt, but technological exposure alone does not determine uptake \parencite{Ransom2026ARegions}. Second, adoption is unequal across workers within the same occupation, with younger, more educated, and male workers consistently more likely to report use, raising concerns that AI may amplify existing labour market inequalities \parencite{Henseke2025WhatAdoption,Humlum2025TheWorkers, DellAcqua2026NavigatingQuality}. Third, adoption varies markedly across countries, yet the institutional and structural factors that explain this variation remain poorly understood \parencite{Bick2026MindU.S.}.

Beyond who adopts, a second open question concerns what AI does to work itself. Experimental evidence indicates that generative AI use raises individual productivity \parencite{Brynjolfsson2025GenAI,Noy2023ExperimentalIntelligence, DellAcqua2026NavigatingQuality}, and early evidence from digital freelance platforms points to declining demand for AI-exposed activities \parencite{Demirci2025WhoPlatforms,Teutloff2025WinnersDemand}. Yet evidence for generative AI's broader economic and labour-market effects has been mixed \parencite{Brynjolfsson2025CanariesTo,Humlum2025LaborMarket,KleinTeeselink2025GenerativeKingdom}. A task-based framework predicts that the margin of adjustment is neither the job nor the worker but the bundle of tasks workers perform: AI displaces some tasks, creates others, and reallocates time across the remaining activities \parencite{Acemoglu2022ArtificialVacancies}. Within this framework, generative AI's labour-market effects differ from those of earlier digital technologies in terms of what jobs, tasks, and workers stand to benefit and which face displacement pressure \parencite{Cnossen2025TasksTechnologies, Bloom2025ArtificialPremium}. However, whether these task-level adjustments are underway in European workplaces, and whether they can be distinguished from the sorting of AI users in already-changing jobs, remains largely untested.

Addressing these gaps requires data that spans multiple countries, covers a representative sample of the employed labour force, and captures the full range of jobs. Existing studies are predominantly single-country, focus on a specific labour market segment, or draw data from online access panels, leaving open two questions: how European labour market institutions, digital infrastructure, and workforce composition shape the adoption of generative AI, and whether early adoption has begun to reshape the task content of jobs.

This paper fills that gap using the 2024 European Working Conditions Survey (EWCS), administered face-to-face to approximately 36{,}600 workers across 35 European countries, covering the 27 EU Member States (EU27), Norway, Switzerland, and several candidate or potential candidate countries \parencite{eurofound_2026_ewcs_overview}. The paper's contributions are as follows:

\begin{enumerate}
    \item It documents the cross-national distribution of generative AI adoption across 35 European countries, the widest such evidence base to date, and benchmark worker-reported use against independent enterprise-level measures from Eurostat. Across Europe, 12\% of workers used generative AI for their job, but with country differences ranging from under three percent to approximately a quarter of the employed workforce.
    \item It estimates the occupational exposure-to-adoption gradient using a task-based measure of generative-AI exposure \parencite{Eloundou2024GPTs, Henseke2025HowIndex} and examines how its steepness varies across workers, jobs, and organisational settings. The findings show that occupational exposure is a strong predictor of adoption, but its strength depends on enabling conditions: tertiary education, abstract job content, and employee say in workplace decisions steepen the exposure--adoption gradient, while workplace size and task routineness do not. The pattern is consistent with individual skills and job content shaping adoption alongside, rather than exclusively through, organisational policy.
    \item It documents a gender gap that is concentrated in the most exposed occupations, extending single-country evidence \parencite[e.g.,][]{Aldasoro2024TheGap,Humlum2025TheWorkers} to a pan-European context.  
    \item It exploits the 35-country design to examine which institutional and structural factors predict the steepness of the exposure--adoption gradient at the country level, identifying digital intensity, measured by the share of workers regularly using computers, training provision, and the abstract task content of the national job stock as relevant correlates.
    \item It examines whether early adoption has begun to reshape the task content of jobs. Using technology-generic worker-reported task-change measures and a cross-country shift-share design, no clearly detectable effect is observed of early AI adoption on perceived task displacement or task creation.
\end{enumerate}

This study adds to a growing body of work that examines the diffusion of AI at work and its emerging effects on jobs. \textcite{Liu2026WhoAI} compare generative AI engagement across countries in 2024 using web traffic and Google Trends data, finding systematic cross-country differences in AI diffusion related to the share of youth population, digital infrastructure, English fluency, foreign direct investment inflows, services' share of GDP, and human capital. However, the paper cannot identify who uses AI at work, nor the consequences of adoption for job content. \textcite{Humlum2025LaborMarket} document widespread adoption of generative AI in 11 highly exposed Danish occupations, with users more likely than non-users to report a reorganisation of work around new tasks and a reallocation of time away from AI-augmented activities, though these micro-level disruptions have not yet broken through to earnings or hours. Unlike them, the current study draws instead on nationally representative samples of the entire employed workforce aged 16--74 years. Like them, I examine worker-reported task restructuring in response to new technologies. I do so by exploiting the cross-country variation in AI adoption within similar occupations in the EWCS to separate causal effects from the sorting of workers in already-changing jobs into adoption. Closest to the current study is \textcite{Bick2026MindU.S.}, who also use survey evidence to examine generative AI adoption at the worker level in Europe. Three differences are worth noting. First, scope: the EWCS covers 35 countries, supporting a second-stage analysis of country-level institutional correlates. Second, focus: their central concern is the U.S.-Europe adoption gap, which they show is largely accounted for by differences in demographics, firm composition, and employer encouragement of AI use. Mine is the translation of occupational AI exposure into individual uptake, and the individual, job, organisational, and institutional conditions that moderate it. Third, interpretation: they identify employer encouragement as the dominant predictor of cross-country differences; the findings presented here suggest that individual skills, abstract task content, and employee say in workplace decisions shape adoption, consistent with the informal, worker-led adoption patterns documented by \textcite{Arntz2026LowBarriers} in German linked employer-employee data.       

The remainder of the paper proceeds as follows. Section~\ref{sec:background} reviews the theoretical framework and empirical literature. Section~\ref{sec:data} describes the data and variables. Section~\ref{sec:methods} outlines the analytical strategy. Section~\ref{sec:results} presents results. Section~\ref{sec:discussion} discusses implications, limitations, and concludes.

\section{Background}
\label{sec:background}

\subsection{Occupational Exposure and Technology Adoption}
\label{ss:task_framework}
Research on the labour market effects of AI suggests that Large Language Models (LLMs) and related generative AI tools extend the automation frontier to tasks requiring language comprehension, text and code generation, and information synthesis, domains that earlier waves of automation left largely untouched \parencite{Acemoglu2022ArtificialVacancies,Felten2023OccupationalAI, Handa2025WhichConversations}. Unlike robotic automation and many tangible digital technologies, which are concentrated in manufacturing and routine manual and cognitive work, generative AI susceptibility is particularly elevated in higher-paying, white-collar, professional occupations such as software engineers and data scientists \parencite{Acemoglu2020RobotsMarkets,Prytkova2024TheTechnologies,Eloundou2024GPTs}.

The task-based framework \parencite[e.g.,][]{Hampole2025ArtificialMarket} predicts that the economic incentives to use AI will increase with occupations' AI task-susceptibility. Highly exposed jobs are precisely those where AI use is most likely to significantly reduce task completion time across a range of work activities. Empirically, this prediction is supported: \textcite{Bick2024TheAI} shows AI adoption ranged from nearly half of all U.S. workers in computer/maths occupations to one in eight in personal services. Similarly, in Denmark and the UK, workers in occupations with high exposure to large language models report substantially higher AI adoption rates \parencite{Henseke2025HowIndex, Humlum2025TheWorkers}. However, the mapping from technical exposure to technological adoption is far from mechanical \parencite{Brynjolfsson2021JCurves, Bresnahan2002ICT}. Across occupations with ostensibly similar AI susceptibility, adoption rates vary widely, indicating that worker and workplace characteristics moderate the translation of potential into actual use \parencite{McElheran2024AIWhere}.

\subsection{Worker and Workplace Moderators of Adoption}

What determines whether individual workers in similarly exposed occupations actually take up the technology? A growing body of evidence points to three classes of moderators: worker human capital, job content, and organisational context. A parallel literature documents a persistent gender gap in adoption that operates within each of these categories.

\textbf{Human capital.} Education, age, and training proxy for the stock of general skills workers bring to the adoption decision. The decision to use generative AI involves non-trivial learning and deployment costs, such as learning to formulate effective prompts, evaluating outputs, and integrating the tool into existing workflows, that accrue to the individual regardless of the tool's nominal user-friendliness. Workers with more formal education, and those engaged in recent skill-updating activities, may be better positioned to absorb these costs. Consistent with this, \textcite{Humlum2025TheWorkers} find that college-educated workers in exposed Danish occupations are substantially more likely to adopt ChatGPT, and that among non-adopters a perceived need for training is the most commonly cited barrier. \textcite{Arntz2026LowBarriers} similarly report that education and job complexity are the strongest individual-level predictors of AI use in Germany.

By contrast, age is a consistent negative correlate of generative AI adoption in worker-level studies \parencite{Bick2026MindU.S.,Humlum2025TheWorkers,Arntz2026LowBarriers}, consistent with patterns observed in earlier technological transitions, including computerisation, internet, and cloud adoption \parencite[e.g.,][]{Friedberg2003TheUse}. We thus include age as a demographic control throughout the subsequent analyses. 

\textbf{Task content.} Within occupations of similar AI susceptibility, what workers actually do and how much control they exercise over their task content vary considerably, shaping both the opportunity and the incentive to adopt. Jobs with high levels of non-routine cognitive task content, i.e., those involving abstract reasoning including analysis, synthesis, problem-solving, reading and writing, offer entry points for generative AI augmentation, because LLM capabilities augment precisely these domains \parencite{Eloundou2024GPTs}. Routine tasks present a more ambiguous case. On the one hand, short, repetitive work may be readily automated by LLM-based tools \parencite{Brynjolfsson2018WhatEconomy}; on the other hand, highly routinised jobs may have been under automation pressure from digital technologies predating and going beyond generative AI \parencite{Cnossen2025TasksTechnologies}. Moreover, workers in these jobs often operate within rigid procedural templates that leave little scope for self-initiated experimentation. Priors thus point in both directions, and whether routineness operates as a barrier or an accelerator to AI adoption is an empirical question.

A long-standing literature in organisational economics and the sociology of work identifies worker discretion, or job autonomy, as a precursor to successful technology adoption: workers with control over how they allocate time and structure tasks can experiment with new tools, adapt them to specific work problems, and integrate them into established routines without managerial direction \parencite{Bresnahan2002ICT,Green2012EmployeeAnalysis}. Conversely, where job structure is tightly prescribed, whether by management, by production processes, or by regulatory constraints, the scope for bottom-up adoption narrows.

\textbf{Organisational context and worker involvement.} Beyond what individual workers do and how much control they exercise over their own tasks, the organisation itself shapes the conditiopn for AI use. The workplace acts as both a potential gatekeeper through explicit restrictions and as a potential accelerator through training provision, tool procurement, and legitimation of use. Recent evidence suggests both channels are operative. \textcite{Humlum2025TheWorkers} show that employer encouragement, including the provision of enterprise AI tools and active promotion of their use, substantially raises adoption and intensity of use, while employer restrictions remain a leading barrier among non-adopters. \textcite{Bick2026MindU.S.} provide complementary cross-national evidence, showing that the quality of human resource management is associated with higher rates of worker AI adoption, operating through greater encouragement in better-managed organisations, including access to AI tools, training, and formal endorsement of use. Beyond formal employer practices, the organisational conditions under which workers themselves shape how new tools are introduced into their jobs may also matter. A relevant feature in this regard is whether the workplace empowers workers to have a say in how their job and their workplace are organised through high-involvement management practices \parencite{Felstead2010EmployeeAnalysis,AppelbaumEileen2001HPWp}. Direct institutional pathways for participation in decision-making empower workers to contribute to the design, deployment, and continued improvement of new workplace technologies, potentially helping to mitigate adverse outcomes for individual jobs and to redistribute productivity gains \parencite{Berg2023RisksChallenge}. Involving workers in workplace decision-making thus raises the likelihood that new technologies are trusted, taken up, and used productively after implementation \parencite{Milanez2023TheImplementation}. Workplace size may additionally matter through resource constraints, scale economies in IT infrastructure, and internal knowledge-sharing \parencite{McElheran2024AIWhere}. Whether these organisational features predict worker AI adoption in practice is, however, theoretically ambiguous: generative AI's low marginal costs of deployment and nominal ease of use \parencite{Arntz2026LowBarriers} mean that workers can, in principle, access AI tools independently of employer provision, potentially attenuating the role of workplace structures in shaping who uses AI at work.

\textbf{Gender.} Cutting across these categories, a consistent gender gap has emerged in early generative AI adoption. \textcite{Liu2026WhoAI} show that only around one-third of ChatGPT users globally are female. In the UK, \textcite{Henseke2025WhatAdoption} document a six percentage point gender gap. \textcite{Humlum2025TheWorkers} find that women in 11 exposed Danish occupations are 16 percentage points less likely than men to have used generative AI for work, and that the gap persists when comparing coworkers within the same workplace and task mix. Women were more likely to cite a perceived need for training and to express greater uncertainty about the tool's usefulness, while men's non-adoption is more often attributed to employer restrictions. Gender thus operates as a moderator of the exposure--adoption link rather than as a function of occupational sorting, and we include it as one conditioning factor.

\subsection{Country-Level Variation and Institutional Context}

Historically, technology diffusion has been shaped by national differences in digital infrastructure, human capital endowments, and labour market institutions \parencite{Comin2004Cross-countryFacts, Comin2010AnDiffusion}. Cross-country studies of earlier general-purpose technologies show that adoption lags vary by decades even among advanced economies, with the speed of diffusion strongly predicted by prior adoption of predecessor technologies and by the absorptive capacity of the workforce \parencite{Comin2010AnDiffusion}. For generative AI, similar mechanisms are likely to operate \parencite[e.g.,][]{Brunetti2025TechnologicalFirms}. LLM-based tools presuppose routine computer use: where digital hardware is not yet embedded in daily workflows, the infrastructure for AI adoption is absent, regardless of occupational exposure.
Beyond digital infrastructure, the workforce's human capital composition may shape how effectively exposure translates into adoption. A more skilled workforce is better able to evaluate and implement new technologies, an absorptive capacity argument that extends to generative AI, where effective use requires human judgement about when and how to deploy the tool \parencite{Keller2004InternationalDiffusion,DellAcqua2026NavigatingQuality}. Countries investing more heavily in workplace training systems may generate steeper exposure--adoption gradients, not because individual training is AI-specific, but because it maintains a workforce capable of integrating new tools as they arise. Organisational practices also vary systematically: \textcite{Bick2026MindU.S.} show that cross-country differences in human resource management practices account for the bulk of the U.S.--Europe adoption gap. Finally, aggregate productivity correlates with economic structures that are more technology and knowledge-intensive, potentially facilitating the adoption of AI tools. Distinguishing these compositional from institutional channels is a central objective of our following multi-level analysis across 35 European countries.

\subsection{Task Restructuring as a Margin of Adjustment}

If aggregate earnings and hours effects have so far been muted, where should we expect early signs of AI's labour market impact to appear? The task-based framework suggests the relevant margin of adjustment is the composition of work within jobs rather than the creation or destruction of jobs themselves. Under a task-level automation model, a worker who adopts a technology that substitutes for a subset of their activities should, in the short run, either shed the automated tasks and reallocate time to other activities, or absorb time savings into a reorganised task portfolio that includes new AI-related activities, such as prompt formulation, output evaluation and iteration, or the integration of generated content into downstream workflows \parencite{Hampole2025ArtificialMarket}. In principle, both displacement and creation should be observable to workers themselves, even before they show up in aggregate labour market statistics.

Direct evidence on whether AI adoption is restructuring tasks remains limited. Pre-generative-AI evidence indicates potentially heterogeneous within-job effects of AI use. Using a Danish employee survey, \textcite{Holm2022TheDenmark} find that AI is associated with skill upgrading in high-skilled occupations but, where AI is used to direct rather than inform workers, with tighter pace constraints and reduced 
autonomy. Turning to generative AI, recent evidence from Denmark indicates approximately eight percent of chatbot users report entirely new tasks arising from AI use, rising to 17\% in workplaces with active employer initiatives, and time savings are predominantly reallocated to other work rather than leisure \parencite{Humlum2025LaborMarket}. However, adopters likely differ systematically from non-adopters on characteristics that predict task change: workers in occupations undergoing broader digitalisation, in firms investing in new work processes, or in roles where task discretion is already high may adopt AI and report restructuring for reasons unrelated to adoption itself. In contrast, experimental evidence indicates that access to Microsoft 365 Copilot, a generative AI tool integrated into Microsoft Office, reduced average time on email but did not shift the quantity or composition of workers' tasks \parencite{NBERw33795}. Distinguishing the causal effect of AI adoption on task composition from the selection of already-changing jobs into adoption requires variation in exposure that is plausibly exogenous to unobserved drivers of task change: a problem we address in Section~\ref{sec:impacts} using a shift-share design.

\section{Data and Measures}
\label{sec:data}

\subsection{European Working Conditions Survey 2024}

Our primary data source is the eighth European Working Conditions Survey (EWCS), conducted by Eurofound in 2024. The EWCS is a cross-sectional survey of employed individuals across 35 European countries (EU member states plus several candidate and associated countries), administered via computer-assisted personal interviews to a random probability sample of the working-age population. Interviews lasted on average 45 minutes. The 2024 wave covers approximately 36{,}600 face-to-face respondents in work, both employees and the self-employed. Fieldwork was conducted predominantly in the first half of 2024 \parencite{eurofound_2026_ewcs2024}.

Analytical sample sizes vary across specifications due to sample restrictions (e.g., the analysis of job task restructuring in Section~\ref{sec:impacts} is limited to workers who use a computer at work) and item non-response on explanatory variables; exact sample sizes are reported in each table. All individual-level estimates are weighted using the EWCS calibrated survey weight, which combines design weights reflecting differential selection probabilities with non-response adjustments and calibration to known national labour force statistics. Standard errors in micro-level estimations are clustered at the country-by-two-digit-occupation level to reflect the structure of the occupational exposure measure.

\subsection{Generative AI Adoption}

Measuring workers' use of generative AI is not straightforward, and no consensus measure has emerged in the literature. The EWCS 2024 asks respondents: \emph{``At work, do you use artificial intelligence that simplifies complex mental tasks or makes recommendations (e.g.\ ChatGPT, LLAMA, DALL-E, Midjourney, Jasper)?''} ($1 = $ Yes, $2 = $ No). The question stem is broader than generative AI: ``simplifies complex mental tasks or makes recommendations'' could encompass AI-powered recommendation systems or decision-support tools that workers interact with directly, not only generative models. Moreover, the measure captures only the extensive margin of use, with no information on frequency or depth of integration. It will also capture only AI tools that are salient to workers, not AI embedded in backend processes that workers are unaware of. However, the named examples anchor respondent interpretation firmly toward large language models and related generative tools, and the ``at work'' qualifier excludes purely personal use. Given available cross-country survey instruments, this item provides the most direct identification of workers who use generative AI tools in a workplace context. Despite the differences in scope between worker-reported and organisational AI use, a comparison of EWCS adoption rates with the share of enterprises using AI in Europe across industry sections (NACE Level~1) and countries shows strong convergent validity, with a precision-weighted correlation coefficient of $r = 0.86$ ($p < .001$; $N = 303$ industry-by-country cells, see Figure~\ref{app:fig:validation} for the corresponding scatter plot).\footnote{ Enterprise-level AI adoption is measured as the share of enterprises using at least one AI technology, drawn from the Eurostat Community Survey on ICT Usage and E-Commerce in Enterprise Survey 2024 (doi:~\href{https://doi.org/10.2908/ISOC_EB_AIN2}{10.2908/ISOC\_EB\_AIN2})}

\subsection{Outcome Measures: Task Displacement and Creation}

To examine early consequences of AI adoption, we use two binary items from the EWCS technology impact battery: (a) ``Technology has removed some of your tasks'' (\emph{Task Displacement} and (b) ``Technology has created new tasks for you'' (\emph{Task Creation}, each recoded as one if the respondent indicated the statements applies to a large or to some extent. \emph{Task Restructuring} is the joint set of displacement and creation. The items enable an assessment of the degree to which AI adoption predicts worker-reported technology-related task restructuring. 

\subsection{Occupational Generative AI Susceptibility}
\label{ss:gaisi}

To measure the potential of each worker's job to be assisted or disrupted by generative AI, we use the Generative AI Susceptibility Index (GAISI) developed by \textcite{Henseke2025HowIndex}. GAISI is built by applying the exposure rubric of \textcite{Eloundou2024GPTs}, which classifies tasks by whether large language models can meaningfully reduce the time required to complete them, either directly or via LLM-integrated software, to the task items in the OECD's 2023 Survey of Adult Skills (PIAAC), using large language models as raters \parencite{OECD2024ReadersCompanionPIAAC2023}.\footnote{PIAAC: OECD's Programme for the International Assessment of Adult Competencies.} Task-level exposure ratings are combined with PIAAC task-frequency weights to generate a worker-level exposure score. Jobs in the top quintile of the resulting distribution are flagged as highly exposed. GAISI at the two-digit ISCO level is then the share of workers within each occupation who fall into this highly exposed group. Higher values, therefore, indicate a greater concentration of highly AI-susceptible jobs within an occupation. According to the index, information and communications technology professionals and technicians, followed by legal, social and cultural professionals, are the most AI-exposed occupation groups, while routine occupations such as refuse workers or food preparation assistants have the lowest exposure. \textcite{Henseke2025HowIndex} document that this LLM-based rating procedure yields reliable and valid measures of occupational exposure across alternative raters, prompts, and weighting choices.

GAISI scores are merged onto individual EWCS respondents via their two-digit ISCO occupation code and standardised to mean zero and unit standard deviation across the analytical sample. The index thus varies across occupations but is common within each two-digit occupation--country cell. Two features of the measure are worth noting. First, GAISI captures technological \emph{exposure}, not adoption: it describes the share of jobs within an occupation whose task content current LLMs could plausibly augment, independent of whether workers or their employers are taking up the technology. Second, the index is an occupation-level constant applied to all 35 EWCS countries, including several not represented in PIAAC 2023; it thus assumes that the task content of a given two-digit ISCO occupation is broadly comparable across European labour markets. As a robustness check, Appendix~\ref{app:emmr} replicates the main exposure--adoption slope estimates using the human-rater exposure scores of \textcite{Eloundou2024GPTs} mapped directly to two-digit ISCO. The substantive pattern is unchanged, though the measure yields slightly less precise predictions of AI uptake than our preferred measure (Table~\ref{tab:emmr_robust}).

\subsection{Individual-Level Moderators and Controls}
\label{ss:ind_vars}

The moderator analysis in Section~\ref{ss:moderators} tests the individual, job, and organisational channels discussed in Section~\ref{sec:background}. We describe the corresponding measures below, grouped by theoretical construct.

\paragraph{Human capital.} \emph{Tertiary educational attainment} is a binary indicator equal to one for respondents who completed a short-cycle tertiary programme or higher, derived from the highest reported qualification classified to the International Standard Classification of Education (ISCED). \emph{Workplace training} is a binary indicator equal to one if the respondent received employer-provided or self-financed training in the twelve months preceding the interview. Age and job tenure, which proxy for workforce experience and workplace attachment, enter as controls.

\paragraph{Job content.} Two indices capture heterogeneity in job content within occupations. \emph{Abstract tasks} is the first factor from a principal components factor analysis with varimax rotation of eight EWCS task-content items, loading on problem-solving, learning, complex task execution, use of foreign languages, and difficult decisions. \emph{Routine tasks} is the second factor from the same analysis, loading on monotonous and repetitive task items. Both factors are standardised to mean zero and unit variance. Full loadings are reported in Appendix~Table~\ref{tab:factor}. \emph{Task discretion} is a standardised index capturing control over how work is performed at the task level: the order of tasks, methods of work, and speed of work 
($\alpha = 0.85$). \emph{Frequent computer use} is a binary indicator equal to one for respondents who report using a computer at work almost all or all of the time, and zero otherwise. It captures the extensive margin of digital work and serves as a precondition for generative AI use. 

\paragraph{Organisational context.} \emph{Individual organisational influence} measures employee voice in decisions affecting the wider organisation, conceptually distinct from task-level discretion. Drawing on 
\textcite{Gallie2020EmployeeDevelopment}, the index averages four items asking how often the respondent is consulted before objectives are set, involved in improving work processes, able to influence decisions important for their work, and able to apply their own ideas ($\alpha = 0.79$; five-point scale reversed so higher values indicate greater influence, then z-standardised in the pooled sample of employees). \emph{Large workplace} is a binary indicator for employment in workplaces with more than 100 workers, entered as a moderator alongside the full workplace-size categorical (below) to test scale effects on adoption.

\paragraph{Controls.} All individual-level specifications include gender, age in five-year bands, a binary indicator for foreign-born (country of birth differs from country of residence), self-employment status, part-time employment, job tenure, and workplace size as a five-category variable. Industry is controlled for via two-digit NACE fixed effects in specifications so noted.

\subsection{Country-Level Indicators}
\label{ss:country_vars}

To examine institutional and structural country-level correlates of the occupational exposure--adoption gradient, we construct a set of aggregate measures from the EWCS and external sources. Individual responses are aggregated to the country level using grossing weights to produce population-representative averages. \emph{Digital intensity} is the share of workers reporting frequent computer use, capturing the infrastructure prerequisite for generative AI adoption. \emph{Workplace training provision} is the share of workers who received employer-provided or self-financed training in the past twelve months, proxying the capacity of national training systems to sustain workforce absorptive capacity. Mean \emph{abstract task intensity} captures the cognitive intensity of the national job stock. The \emph{tertiary education share} and the \emph{self-employment share} measure formal human capital endowments and the prevalence of independent work outside formal employer-employee relationships, respectively. Finally, to capture underlying economic productivity differences, we import log GDP per worker in 2019 from the Penn World Table (version 11) \parencite{feenstra_inklaar_timmer_2015}; the pre-pandemic, pre-ChatGPT reference year isolates structural productivity from any AI-induced shifts. In country-level regressions, observations are precision-weighted by EWCS cell sample counts rather than grossing weights.

\section{Analytical Strategy}
\label{sec:methods}

The analysis proceeds in four steps. The first three characterise the exposure--adoption slope, which expresses how well occupational potential correlates with individual uptake, and the worker, job, organisational, and country-level characteristics that condition it. The final step moves to identification, using an aggregate shift-share design to examine whether early adoption has begun to reshape the task content of jobs. All specifications are linear probability models, which handle binary outcomes with high-dimensional fixed effects transparently and yield coefficients directly interpretable as percentage-point marginal effects. Analytical samples are the broadest available for each specification and are narrowed only where conceptually meaningful: the moderation analysis in Step 2 restricts to employees because organisational factors matter only for their adoption, and the task-restructuring analysis in Step 4 restricts to computer users, the population ``at risk'' of generative AI adoption.

\paragraph{Step 1: Pooled exposure-to-adoption gradient.} I estimate the relationship between occupational exposure and individual AI adoption using a sequence of progressively controlled specifications:

\begin{equation}
\label{eq:pooled}
\mathrm{AI}_{ijc} = \beta \cdot \mathrm{GAISI}_j + X'_{ijc}\bm{\gamma} + 
\lambda_c + \epsilon_{ijc},
\end{equation}

where $i$ denotes the worker, $j$ the two-digit occupation, and $c$ the country. $X$ is a vector of individual, job, and organisational characteristics that varies across specifications; $\lambda_c$ are country fixed effects; and standard errors are clustered at the country-by-occupation level. Three specifications are reported: demographics only, adding human capital and computer use, and further adding employment relations, organisational characteristics and industry fixed effects. The purpose of this sequence is decompositional: comparing $\hat{\beta}$ across columns identifies how much of the raw occupation exposure--adoption association reflects skill composition, how much reflects sorting of skilled workers into digital-intensive industries, and how much persists as a within-industry, within-country gradient net of observable worker and workplace characteristics.

A parallel specification examines the gender gap in adoption. We decompose the raw gap using a similar progressive-control sequence, adding demographics, human capital, job content, and organisational characteristics in turn, to identify which compositional factors account for men's higher adoption. We then replace the uniform male indicator with interactions between gender and exposure quintiles to test whether the gap concentrates at specific levels rather than applying uniformly across the distribution. 

\paragraph{Step 2: Moderation analysis.} Having established gender gaps and the pooled occupational exposure--adoption gradient, we turn to the question of what conditions its slope. The slope is the central quantity of interest because it conceptually measures the rate at which technological potential translates into individual uptake: a steep slope implies that workers in exposed occupations are acting on that exposure, while a flat slope implies that exposure alone is insufficient to generate adoption. Identifying what steepens or flattens the slope thus identifies the enabling conditions through which exposure becomes adoption.

We estimate interaction models of the form:

\begin{equation}
\label{eq:moderation}
\mathrm{AI}_{ijc} = \beta_1 \mathrm{GAISI}_j + \beta_2 M_{ijc} + \beta_3 
(\mathrm{GAISI}_j \times M_{ijc}) + X'_{ijc}\bm{\gamma} + \lambda_c + 
\epsilon_{ijc},
\end{equation}

where $M$ is each moderator in turn. All moderation models are estimated on a common employee sample (determined by the full joint model with all moderators) and include country, two-digit occupation, industry, education, and age--sex fixed effects. Because GAISI varies only at the two-digit occupation level and all specifications include occupation fixed effects, the GAISI main effect is absorbed by the occupation FE; the interaction $\beta_3$ is identified from within--occupation variation between workers who share the same occupational exposure but differ in their moderator value. This within-occupation identification accounts for occupation sorting. In other words, it rules out that a moderator appears to steepen the slope because workers with high moderator values (e.g., the tertiary-educated) sort into more exposed occupations. The interaction coefficient captures whether the moderator steepens or flattens the exposure--adoption slope conditional on observed worker, job, and workplace characteristics. Each moderator is first entered separately to characterise the profile of workers who convert exposure into adoption, and then jointly to identify which dimensions carry independent predictive power after accounting for their intercorrelation. 

\paragraph{Step 3: Country-level second stage.} Country-specific GAISI slopes $\hat{\beta}_{1c}$ are recovered from a fully interacted version of Equation~(\ref{eq:pooled}) and used as the dependent variable in a second-stage regression on country-level characteristics:

\begin{equation}
\label{eq:second_stage}
\hat{\beta}_{1c} = \alpha_0 + \alpha_1 Z_c + u_c,
\end{equation}

where $Z_c$ is a country-level characteristic and observations are weighted by precision ($1/\hat{\sigma}^2_{\beta_{1c}}$) to down-weight countries with noisier first-stage estimates, following standard meta-analytic practice. Bivariate models relate the country slope to each correlate in turn; because digital intensity is a logical prerequisite for generative AI use, it is partialled out in the primary specification to isolate correlates that matter beyond fundamental digital infrastructure. The 35-country constraint precludes richer multivariate specifications.

\paragraph{Step 4: Self-reported technology impacts.}
The final step examines whether early AI adoption has begun to reshape the task content of jobs. To separate the causal effect of adoption from the selection of workers in already-changing jobs into adoption, we collapse the data to one-digit ISCO $\times$ country cells, restrict the sample to computer users, and regress cell-level task displacement and creation rates on cell-level AI adoption. Specifications progressively add one-digit occupation and country fixed effects, absorbing tendencies for high-exposure occupations to restructure faster regardless of AI adoption and for common country shocks to affect adoption and task change simultaneously. To address remaining endogeneity in cell-level adoption, we construct a shift-share instrument following \textcite{Borusyak2025AInstruments} that interacts occupational GAISI exposure with the leave-one-out country-average adoption rate, isolating variation in cell-level AI use driven by the interaction between occupational exposure and country-wide adoption intensity. Identification relies on the exogenous-shares interpretation of \textcite{Goldsmith-Pinkham2020BartikHow}: occupational exposure should predict task restructuring only through adoption, conditional on the fixed effects. Cell means are computed using grossing weights and regressions are precision-weighted by cell sample size; standard errors are clustered at the country level. Corresponding individual-level associations between adoption and self-reported task displacement and creation are reported for completeness in Appendix~\ref{app:micro_impacts}.

\section{Results}
\label{sec:results}

\subsection{Patterns of AI Adoption Across Europe}
\label{sec:descriptives}
Before turning to the multivariate analysis, this section explores general patterns of AI uptake at work across Europe. Overall, generative AI adoption stood at 12\% (SD 0.32, $N =$ 36{,}644) across the pooled workforce in the 35 surveyed European countries; 24\% of workers reported technology-induced task displacement, and 34\% reported task creation (N = 36{,}078). In terms of workforce composition, 46\% of workers were female, 53\% were under 45 years old, 11\% were foreign-born, and nearly one in three (32\%) had achieved tertiary qualifications. Looking at employment relations, 15\% of workers were self-employed, close to one in five (19\%) worked part-time, and 45\% had received workplace training in the past year. Of the employed workforce just over half (53\%) used computer all or almost all of the time. 

How does the headline adoption rate of 12\% compare with other estimates of AI diffusion in Europe? Eurostat's 2025 survey of ICT use in households estimates that 32.7\% of 16--74 year-olds across Europe used generative AI platforms in the three months preceding the interview; of these, 46.1\% reported work-related use, implying that roughly 15\% of working-age Europeans used generative AI for work \parencite{EU2026AItech}. A 2024/2025 telephone survey of EU workers suggests that 30\% used ``AI-powered tools'' for work at least once in the 12 months preceding the survey interview \parencite{GonzalezVazquez2025DigitalEurope}, though the broad framing likely captures more than just generative AI. At the firm level, 13.5\% of private-sector enterprises with ten or more employees (excluding agriculture, extractive industries, and finance) in the EU-27 deployed AI technologies in 2024, rising to 20\% in 2025 \parencite{Eurostat2025UsageEurostat}. Restricting EWCS respondents to employees in comparable organisations across the EU-27 yields a worker-level adoption rate of 13.8\% (95\%~CI [12.4, 15.2], $N =$ 9{,}921), closely tracking the enterprise figure. While the measures differ in unit of observation, technology scope, and recall period, the EWCS estimate falls squarely within the range implied by these benchmarks, consistent with the convergent validity documented in Appendix~\ref{app:ss:convergent_validity}.

Figure~\ref{fig:adoption_country} shows generative AI adoption rates at work by country. Adoption varies substantially across the 35 countries in the sample, ranging from under 3\% (Bosnia-Herzegovina) to approximately a quarter (Luxembourg). The distribution exhibits a clear geographical gradient, with higher adoption rates concentrated in Northern and Western Europe than in Southern and Eastern Europe. 

\begin{figure}[H]
    \centering
    \caption{Worker-level AI adoption rates by country.\\
    \label{fig:adoption_country}
    \includegraphics[width=0.9\textwidth]{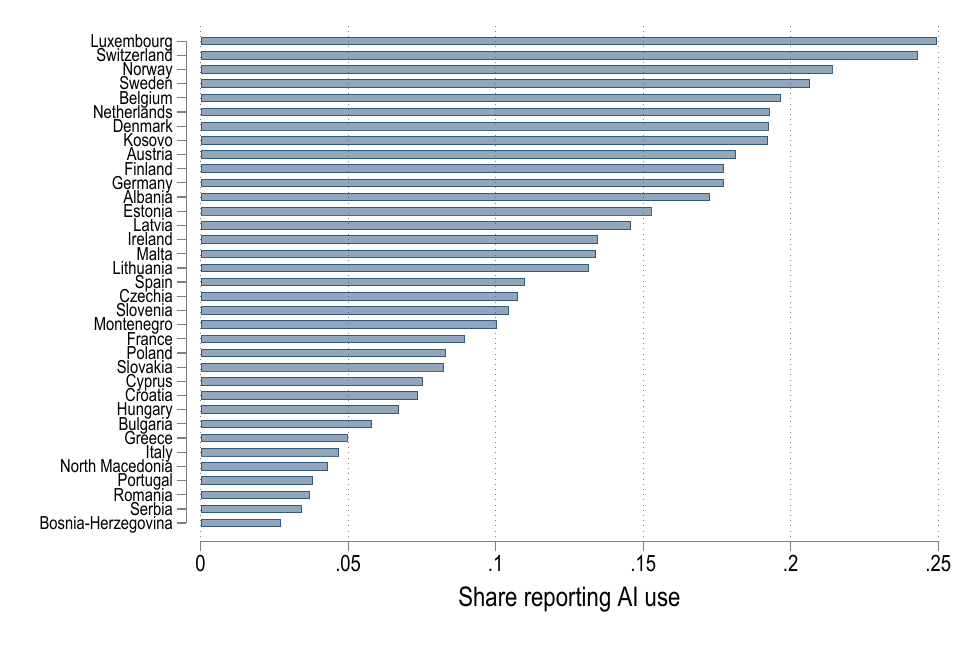}\\
    \footnotesize\textit{Notes:} Weighted share of employed workers reporting use of generative AI at work (EWCS 2024).}
\end{figure}

\begin{figure}[H]
    \centering
    \caption{Gender gap in AI adoption by country (male minus female adoption rate).\\
    \label{fig:gender_gap}
    \includegraphics[width=0.9\textwidth]{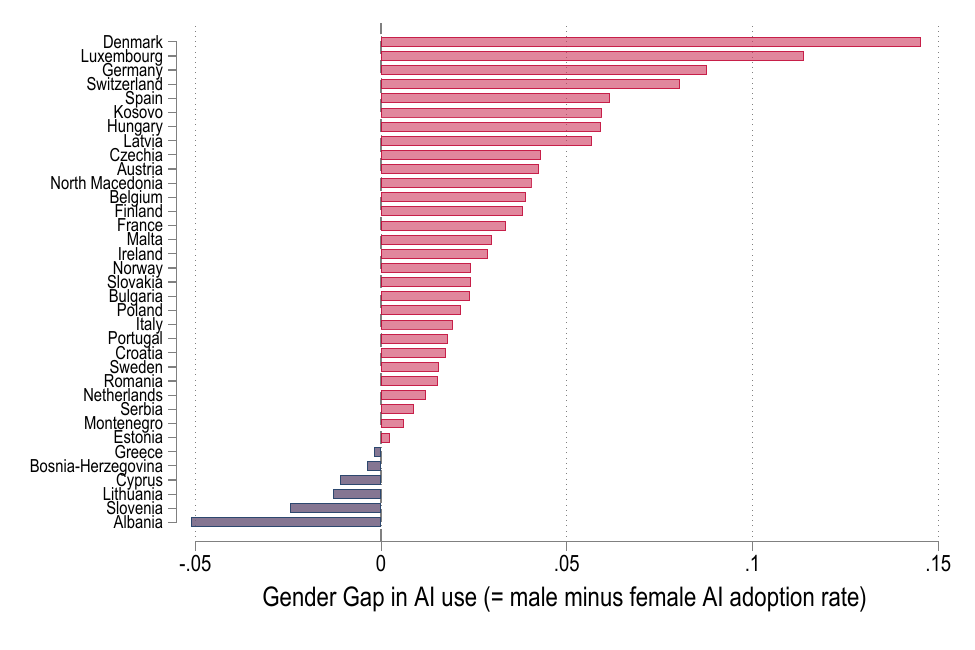}\\
    \footnotesize\textit{Notes:} Difference in weighted AI adoption rates between men and women within each country. Positive values indicate higher male adoption. EWCS 2024.}
\end{figure}

Figure~\ref{fig:gender_gap} presents the male minus female adoption rate by country. The gender gap is positive (men adopt more) in the large majority of countries, though there is considerable variation in its magnitude across the 35-country sample. On average across the surveyed countries, men were 4.1 percentage points more likely to use AI for work than female workers (AI adoption rates of 13.8 vs 9.7 percent), amounting to a gender AI adoption gap of 29.5 percent relative to the male adoption rate.

In all, there has been notable uptake of generative AI across European workplaces within less than two years after the release of ChatGPT~3.5 in November 2022. Adoption rates vary across countries, with Nordic and Western European countries ahead of Southern and Eastern European countries. Gender differences in early adoption are prevalent across most countries.  

\subsection{Occupational Exposure and Adoption}

Having documented the cross-national distribution of AI adoption and the gender gap, the study now examines whether occupational AI exposure organises these patterns.

\begin{figure}[htbp]
    \centering
    \caption{AI adoption by levels of occupational AI susceptibility.}
    \label{fig:ai_by_aie}
    \includegraphics[width=0.9\textwidth]{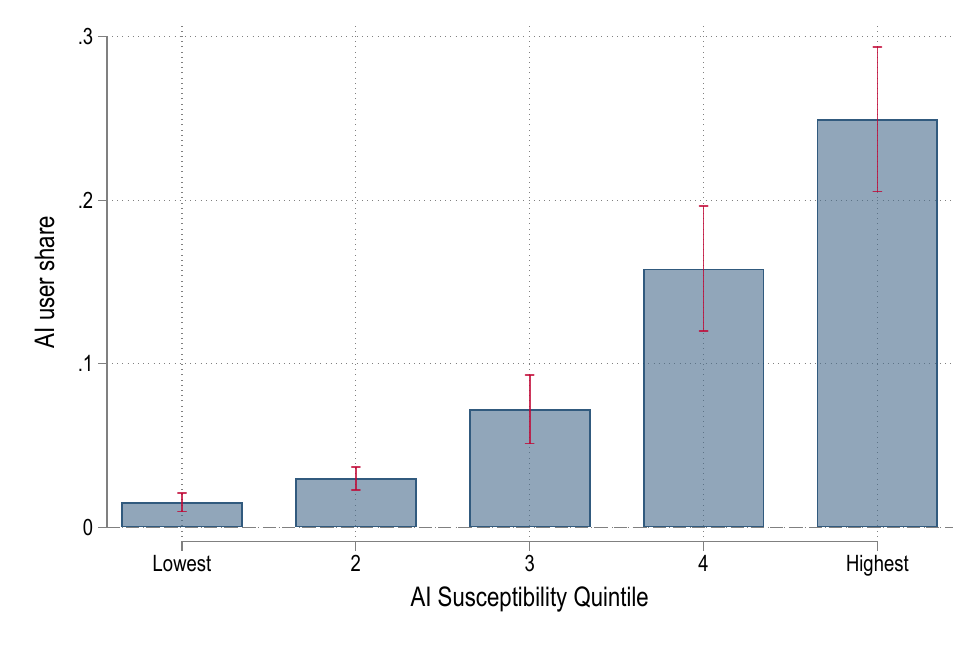}\\
    {\footnotesize\textit{Notes:} Weighted AI adoption rates by quintiles of the Generative AI Susceptibility Index. EWCS 2024.}
\end{figure}

Figure~\ref{fig:ai_by_aie} displays AI adoption rates by quintile of occupational AI susceptibility in the pooled workforce. The gradient is steep: adoption rises from 1.5 percent in the least exposed quintile to nearly a quarter in the most exposed, a gap of 23.4 percentage points. That adoption is non-zero even in the least exposed occupations ($p<0.001$) is consistent with the general-purpose character of generative AI.

Table~\ref{tab:pooled} examines this gradient in a multivariate framework, reporting linear probability models of generative AI use on occupational exposure across three progressively controlled specifications. Column~(1) controls for demographics and country fixed effects; column~(2) adds individual skills and computer use on the job; column~(3) further adds organisational characteristics and industry fixed effects.

\begin{table}[htbp]
\centering
\begin{threeparttable}
\caption{Occupational exposure (GAISI) and generative AI adoption}
\label{tab:pooled}
\small
\begin{tabular*}{0.9\textwidth}{@{\extracolsep{\fill}}l*{3}{c}}
\toprule
 & (1) & (2) & (3) \\
 & Demographics & $+$Skills & $+$Organisation \\
\midrule
\multicolumn{4}{l}{\textit{DV: AI use at work (0/1)}} \\[3pt]
GenAI Susceptibility & 0.101$^{***}$ & 0.066$^{***}$ & 0.049$^{***}$ \\
 & (0.007) & (0.008) & (0.007) \\[3pt]
\midrule
Country FE         & Yes & Yes & Yes \\
Foreign-born, Age$\times$Sex   & Yes & Yes & Yes \\
Education, seniority & No & Yes & Yes \\
Computer use         & No & Yes & Yes \\
Employment status   & No & No & Yes \\
Workplace size     & No & No & Yes \\
Industry FE         & No  & No  & Yes \\
Observations        & 35{,}184 & 35{,}184 & 35{,}184 \\
$R^2$               & 0.147 & 0.182 & 0.229 \\
\bottomrule
\end{tabular*}

\begin{tablenotes}[flushleft]
\footnotesize
\item \textit{Notes:} Linear probability model estimates. The dependent variable is a binary indicator for generative AI use at work. The GenAI Susceptibility Index measures the share of highly exposed tasks in two-digit ISCO occupations, standardised to mean zero and unit standard deviation across occupations. Weighted for design and non-response. Samples restricted to complete cases. Standard errors clustered at country$\times$two-digit occupation (1{,}367 clusters). $^{***}p<0.001$, $^{**}p<0.01$, $^{*}p<0.05$.
\end{tablenotes}
\end{threeparttable}
\end{table}

Occupational AI exposure is a strong and robust predictor of individual adoption. In the baseline specification, a one-standard-deviation increase in GAISI is associated with a 10.1 percentage-point increase of the adoption probability ($p<0.001$), a nearly doubling of the sample mean of 11.9\%. Adding measures of individual skills reduces the coefficient by a third to 6.6 percentage points; further adding organisational controls and industry fixed effects halves it to 4.9 percentage points. The attenuation indicates that part of the raw association reflects skill composition and job sorting into industries and workplace types, but a substantial and precisely estimated exposure gradient persists net of these controls. Results are robust to replacing GAISI with the human-rater exposure scores of \textcite{Eloundou2024GPTs} (Appendix Table~\ref{tab:emmr_robust}).
 
The next section examines how this exposure-adoption gradient varies across workers, examining first the gender gap and then a broader set of individual, job, and organisational moderators.

\subsection{Gender Gap in Generative AI Adoption}
\label{sec:gendergap}

Section~\ref{sec:descriptives} documented a raw gender gap of 4.1 percentage points across 35 European countries. Having established that occupational exposure is a strong predictor of adoption, we now ask how much of this gap reflects differences in human capital, job sorting, and organisational context, and whether men and women convert the same level of occupational exposure into adoption at the same rate. Table~\ref{tab:gender} addresses this through a stepwise decomposition, progressively adding controls to trace the sources of the raw gap, and concluding with GAISI quintile interactions to test whether the gender differential varies with occupational exposure. Compared with the estimation model in the section above, observations drop to 33{,}228 due to higher levels of missingness in task content information.

\begin{table}[htbp]
\centering
\begin{threeparttable}
\caption{Gender gap in generative AI adoption: stepwise decomposition}
\label{tab:gender}
\scriptsize
\begin{tabular*}{\textwidth}{@{\extracolsep{\fill}}l*{5}{c}}
\toprule
 & (1) & (2) & (3) & (4) & (5) \\
 & Demographics & $+$Skills & $+$Occ \& Job & $+$Orga & $+$GAISI$\times$Gender \\
\midrule
\multicolumn{6}{l}{\textit{DV: AI use at work (0/1)}} \\[3pt]
Male     &   $0.043^{***}$ & $0.056^{***}$ & $0.044^{***}$ & $0.040^{***}$ & $-0.005$ \\
           &   (0.011) &  (0.009) & (0.008) & (0.009)  & (0.011) \\ [2pt]
Male $\times$ $2^{nd}$ GAISI Quintile   &       &       &       &       & $0.027$ \\
                     &       &       &       &       & (0.016) \\[2pt]
Male $\times$ $3^{nd}$ GAISI Quintile   &       &       &       &       & $0.041$ \\
                     &       &       &       &       & (0.028) \\[2pt]
Male $\times$ $4^{th}$ GAISI Quintile   &       &       &       &       & $0.033$ \\
                     &       &       &       &       & (0.020) \\[2pt]
Male $\times$ $5^{th}$ GAISI Quintile   &       &       &       &       & $0.080^{***}$ \\
                     &       &       &       &       & (0.018) \\[3pt]                     
\midrule
Country, Age-group, Foreign-born  & Yes & Yes & Yes & Yes & Yes \\
Education, Seniority & No  & Yes & Yes & Yes & Yes \\
Occupation FE, Computer    & No  & No  & Yes & Yes & Yes \\
Abstract, Routineness, Discretion & No & No & Yes & Yes & Yes \\
Employment status, Part-time \\
Industry, Workplace size FE        & No  & No  & No  & Yes & Yes \\
Observations  &  33{,}228 & 33{,}228  & 33{,}228 & 33{,}228 & 33{,}228         \\
$R^2$                 &       0.046         &       0.124         &       0.218         &       0.249         &       0.250         \\
\bottomrule
\end{tabular*}

\begin{tablenotes}[flushleft]
\footnotesize
\item \textit{Notes:} Linear probability model estimates on a common analytical sample. All models include fixed effects for age group, country, and foreign-born status. In columns (3)--(5), occupation fixed effects absorb the GAISI main effect; the GAISI$\times$Female interactions in column (5) are identified from within-occupation variation between male and female workers facing the same occupational AI susceptibility level. Weighted for design and non-response. Samples restricted to complete cases. Standard errors clustered at country$\times$two-digit occupation. $^{***}p<0.001$, $^{**}p<0.01$, $^{*}p<0.05$.
\end{tablenotes}
\end{threeparttable}
\end{table}

Column~(1) documents the gender gap at 4.3 percentage points, conditional on demographics and country fixed effects, close to the raw gap reported in section~\ref{sec:descriptives}. Adding educational attainment and seniority in column~(2) \emph{widens} the gap to 5.6 percentage points. This suppression effect indicates that women's human capital endowments positively predict adoption: once education and experience are held constant, the residual gender penalty is larger than the raw gap suggests. Column~(3) adds occupation fixed effects, computer use, and job content measures (abstract task content, routineness, task discretion), which partially attenuate but do not close the gap (4.4~pp). Column~(4) further adds employment status, industry, and workplace size fixed effects, reducing the gender gap to 4.0 percentage points, marginally below the raw estimate.

Column~(5) replaces the uniform male coefficient with GAISI quintile interactions to test whether the gender gap varies with occupational exposure. The male main effect (now capturing the gap in the least exposed quintile) is small and insignificant ($-0.5$~pp). The interaction terms are small and insignificant in quintiles 2--4, but large and precisely estimated in quintile 5: men in the most AI-exposed occupations are 7.5 percentage points more likely to adopt than women in the same occupations ($=(0.080 - 0.005)*100$, $p<0.001$). In other words, the gender differences are concentrated among the most exposed workers.

This result extends the Danish findings of \textcite{Humlum2025TheWorkers} and other single country results to a 35-country European context. Critically, specifications (3)--(5) absorb occupation fixed effects, so the gap reflects a differential propensity to convert occupational exposure into uptake within the same two-digit occupation, not differences in which occupations men and women hold. Evidence from more recent cross-national evidence from \textcite{Bick2026MindU.S.} finds substantially smaller gender differences in generative AI use by late 2025, raising the possibility that the gap documented here reflects a gendered lag in early adoption rather than a persistent structural disparity. Nonetheless, whether the European gender gap narrows as diffusion progresses and the technology continues to mature, or proves more durable, remains an open question. Section~\ref{ss:moderators} confirms that, at the time of the 2024 EWCS fieldwork, the gender differential holds when estimated jointly with the full set of worker, job, and organisational moderators.

\subsection{From Exposure to Adoption: What Moderates the Link?}
\label{ss:moderators}

The preceding sections established that occupational exposure strongly predicts adoption and that a gender gap persists within occupations. This section asks what job and organisational characteristics condition the translation of exposure into uptake. Since the focus is on organisational antecedents, particularly whether employee involvement in workplace decisions shapes AI adoption, this subsection restricts the sample to employees ($N = 28{,}007$).

Table~\ref{tab:moderators} reports the interaction of GAISI with each candidate moderator. All models include two-digit occupation fixed effects, which absorb the GAISI main effect. The interaction coefficients are identified from within-occupation variation, comparing workers who share the same occupational exposure but differ in their moderator values, rather than from cross-occupation sorting of, e.g., trained workers into high-exposure occupations. The sign and significance of each interaction are informative about direction and relative magnitude, even without a reported baseline slope. Column~(1) estimates each interaction separately, characterising the profile of workers who convert exposure into adoption. Column~(2) enters all moderators jointly, identifying which dimensions carry independent predictive power after accounting for their intercorrelation. The attenuation from column~(1) to column~(2) is expected given the correlation among skill, job content, and organisational variables; we discuss both sets of estimates.

\begin{table}[htbp]
\centering
\begin{threeparttable}
\caption{Moderators of the occupational exposure--adoption gradient (employees, $N = 28{,}007$)}
\label{tab:moderators}
\small
\begin{tabular*}{0.9\textwidth}{@{\extracolsep{\fill}}lcc@{}}
\toprule
 & (1) & (2)  \\
GAISI $\times$ & Separate & Combined \\
\midrule
\ldots Male             & $0.037^{***}$ & $0.031^{***}$  \\
                                  & (0.007) & (0.007) \\[2pt]
\ldots Tertiary educated           & $0.038^{***}$ &  $0.027^{**}$ \\
                                  & (0.010) & (0.009) \\[2pt]  
\ldots Training           & $0.040^{***}$ &  0.027 \\
                                  & (0.013) & (0.014) \\[2pt]
\ldots Abstract         & $0.031^{***}$ & $0.019^{***}$ \\
                                  & (0.004) & (0.005) \\[2pt]
\ldots Routine        & $-0.007$ & $-0.005$ \\
                                  & (0.004) & (0.004) \\[2pt]
\ldots Task Discretion           & $0.017^{***}$ & 0.004 \\
                                  & (0.004) & (0.004) \\[2pt]
\ldots Org.\ influence        & $0.025^{***}$ & $0.016^{***}$ \\
                                  & (0.004) & (0.005) \\[2pt]
\ldots Large workplace (100+)     & $0.024^{**}$ & 0.012 \\
                                  & (0.008) & (0.009) \\[3pt]   
\bottomrule
\end{tabular*}

\begin{tablenotes}[flushleft]
\footnotesize
\item \textit{Notes:} Column~(1) reports results from separate LPMs, each including one GAISI$\times$moderator interaction. Column~(2) reports the joint model with all interactions entered simultaneously. The GAISI main effect is absorbed by two-digit occupation FE in all models; interaction terms are identified from within-occupation variation. Control variables include country, gender$\times$age-group, foreign-born, education, seniority, frequent computer use, part-time, two-digit industry, and workplace size fixed effects. Weighted for design and non-response. Samples restricted to employee cases with complete information. Standard errors clustered at country$\times$two-digit occupation (1{,}345 clusters). Restricted to employees. $^{***}p<0.001$, $^{**}p<0.01$, $^{*}p<0.05$.
\end{tablenotes}
\end{threeparttable}
\end{table}

\paragraph{Gender.}
Consistent with the decomposition in Section~\ref{sec:gendergap}, men exhibit a significantly steeper exposure-adoption gradient ($0.031$, $p<0.001$ in the joint specification). The estimate is robust to controlling for the full set of skill, job content, and organisational moderators, confirming that the gender gap is not attributable to differences in these observable characteristics.

\paragraph{Skills.}
Tertiary education and workplace training both significantly steepen the exposure-adoption gradient when entered separately. In the joint specification, tertiary education retains independent predictive power ($+0.027$, $p<0.01$), while training does not, likely reflecting its overlap with education, jobs' abstract task content and organisational influence. These results are consistent with skill barriers as a first-order friction in converting exposure to uptake \parencite{Humlum2025TheWorkers}.

\paragraph{Task content.}
Abstract task content significantly steepens the gradient in both the separate ($+0.031$) and joint ($+0.019$, $p<0.001$) specifications. Routineness has no significant moderating effect. Task discretion is significant separately ($+0.017$, $p<0.001$) but not jointly, plausibly reflecting its correlation with jobs' task content and individual organisational influence. The pattern indicates that AI adoption concentrates among workers whose tasks demand abstract, cognitive judgement, not merely among those in exposed occupations.

\paragraph{Organisational context.}
Organisational influence, the degree to which employees are consulted on objectives, involved in improving work processes, and able to influence decisions affecting their work, is the most robust organisational moderator. It significantly steepens the exposure--adoption gradient in both the separate ($+0.025$, $p<0.001$) and joint ($+0.016$, $p<0.001$) specifications. By contrast, workplace size is significant only when entered separately and does not survive the joint test. The finding that employees' say in organisational decisions predicts AI uptake beyond individual skills and job content resonates with \textcite{Bick2026MindU.S.}, who find that differences in high-performance work practices account for a substantial share of the AI adoption gap between European and U.S. workers.

\medskip
Three moderators survive the joint specification: tertiary education, abstract task content, and organisational influence, alongside the gender penalty. Training, job autonomy, and workplace size contribute to the profile of the typical early adopter (column~1) but do not carry independent predictive power once the joint model accounts for their shared variance. The pattern suggests that converting occupational AI exposure into adoption depends on the intersection of individual skills, cognitively demanding work, and organisational environments that give workers a say in how their work is organised --- with a persistent gender gap that none of these factors explains.

\subsection{Country-Level Drivers of Adoption Intensity}
\label{ss:countrylevel}

The preceding analysis pooled data from 35 countries, using country fixed effects to control for cross-national differences. The following analysis moves to explore those differences directly. Country of employment remains a jointly significant predictor of AI adoption in the employee sample after controlling for demographics, human capital, occupation and job content, industry, workplace size, and organisational influence ($F = 4.66$, $p < 0.001$), indicating that institutional or structural factors beyond workforce composition shape national adoption patterns.

\begin{figure}[htbp]
    \centering
    \caption{Country AI adoption level and the exposure-adoption slope.}
    \label{fig:slope_vs_level}
    \includegraphics[width=0.9\textwidth]{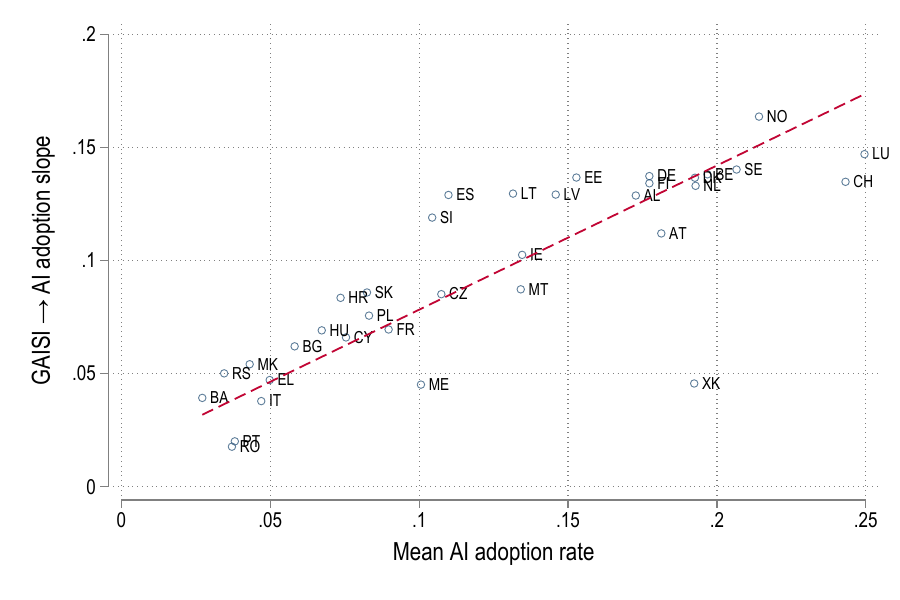}
    
    \begin{minipage}{.9\textwidth}
    \vspace{0.3em}
    \footnotesize
    
    \textit{Notes:} Each point is one country ($N = 35$). The horizontal axis plots the weighted mean AI adoption rate; the vertical axis plots the country-specific marginal effect of GAISI on AI adoption from a fully interacted LPM. Dashed line: OLS fit. Pearson $r = 0.89$
    ($p < 0.001$). EWCS 2024.
    \end{minipage}
\end{figure}

First, occupational exposure significantly predicts adoption in almost all 35 countries at conventional levels of significance, except in Bosnia and Herzegovina, Kosovo and Romania. In other words, occupational exposure shapes adoption nearly everywhere it is taking place; what varies is the rate at which it does so (Figure~\ref{app:fig:gaisi_slopes}). Second, Figure~\ref{fig:slope_vs_level} asks whether countries with higher overall adoption are also countries where the relationship between occupational exposure and uptake is stronger. The relationship is strong and positive ($r = 0.89$): countries with higher mean adoption rates also exhibit steeper GAISI--adoption gradients. This is not mechanical: a uniform upward shift in AI use across all occupations would raise the level without steepening the slope. Rather, the pattern indicates that in high-adoption countries, it is specifically workers in AI-exposed occupations who drive the aggregate adoption rate, consistent with the enabling-conditions interpretation developed in Section~\ref{ss:moderators}. Conversely, in low-adoption countries such as Romania and Portugal, the exposure-adoption gradient is near zero: even workers in highly susceptible occupations rarely adopt. Montenegro and Kosovo exhibit a different pattern: moderate overall adoption but flat slopes, suggesting that what AI use exists is not structured by occupational exposure as measured here. The question that follows is: which country-level characteristics explain this variation?

\begin{figure}[htbp]
    \centering
    \caption{Country-level partial correlates of the exposure--adoption slope, 
conditional on digital intensity.}
    \label{fig:partial_correlates}
    \includegraphics[width=0.9\textwidth]{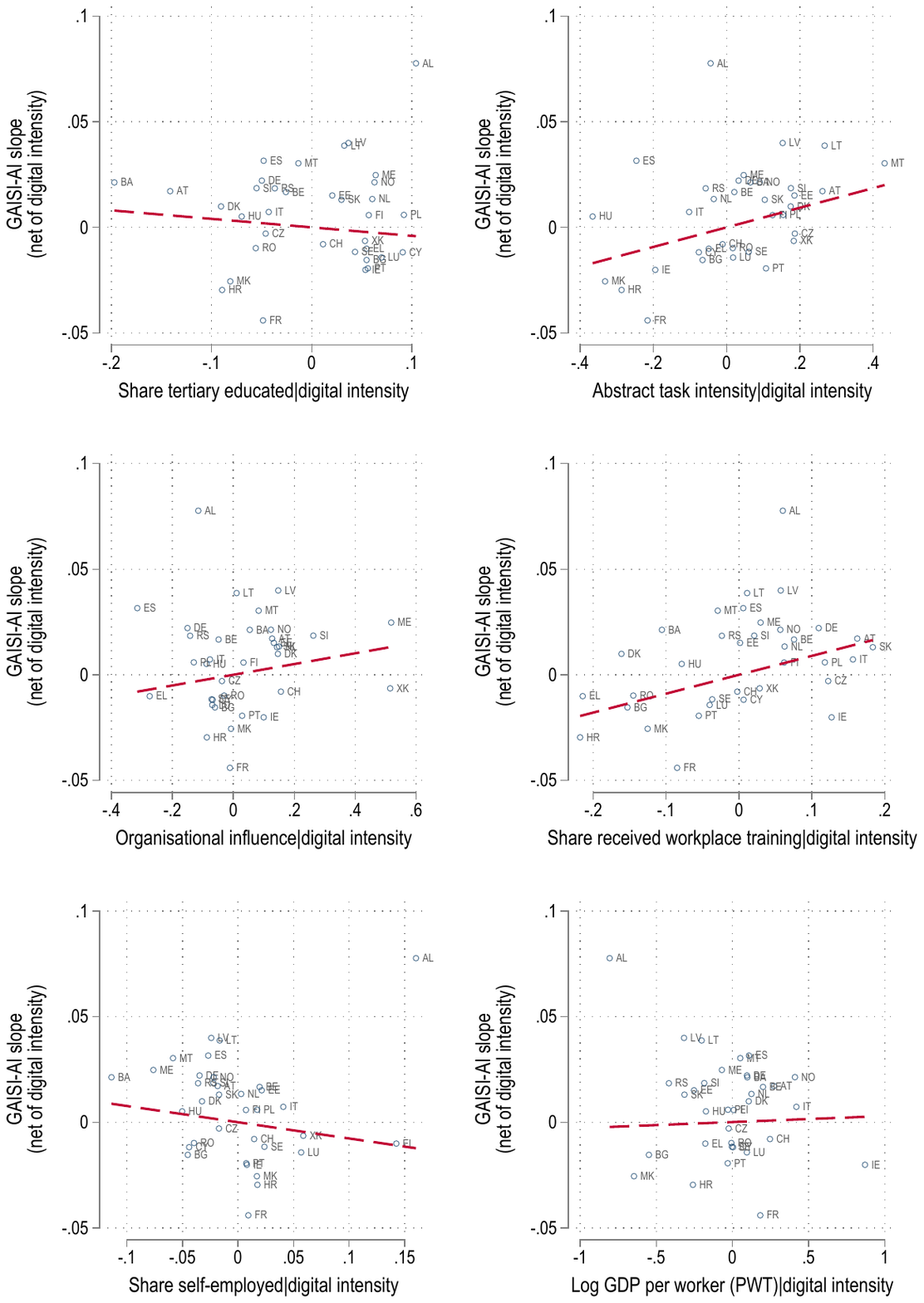}
    
    \begin{minipage}{\textwidth}
    \vspace{0.3em}
    \footnotesize
    \textit{Notes:} Each point represents one country ($N = 35$). Both axes show residuals from precision-weighted regressions on digital intensity (share of workers using a computer almost all or all of the time). The vertical axis is the country-specific GAISI--adoption slope; horizontal axes show country-level workforce and institutional characteristics. Dashed lines: OLS fit. Partial correlations (precision-weighted): tertiary education $r = -0.13$; abstract task intensity $r = 0.41^{*}$; organisational influence $r = 0.19$; training provision $r = 0.49^{**}$; self-employment $r = -0.18$; log GDP per worker $r = 0.04$. EWCS 2024.
    \end{minipage}
\end{figure}

Figure~\ref{fig:partial_correlates} addresses this by exploring six country-level characteristics as partial correlates of the exposure--adoption slope, each conditional on digital intensity --- the share of workers using a computer almost all or all of the time. Digital intensity is a prerequisite for generative AI use and the dominant raw predictor of cross-country slope variation ($r=0.88$); the partial correlations isolate associations \emph{beyond} fundamental digital infrastructure.

Two of the six tested characteristics emerge as significant partial correlates. Training provision, which is the share of workers receiving employer-provided or self-financed training, is the strongest ($r = 0.49$, $p < 0.01$). Countries where workplace training is more prevalent, conditional on digital intensity, exhibit steeper exposure--adoption gradients, suggesting that training systems play a role in enabling workers in exposed occupations to act on their AI adoption opportunities. Abstract task intensity, the average non-routine cognitive content of the national job stock, is the second significant correlate ($r = 0.41$, $p < 0.05$): countries where the average job requires more non-routine cognitive efforts convert occupational exposure into adoption more efficiently.

The remaining four characteristics are not significant partial correlates. Tertiary education is weakly negative ($r = -0.13$), indicating that more educated workforces adopt AI because they use computers, not because of education per se. Organisational influence, one of the robust micro-level moderators in Section~\ref{ss:moderators}, does not aggregate to the country level ($r = 0.19$, $p = 0.29$): variation in employee voice matters within countries but does not differentiate between them, according to our analysis. Self-employment is negative ($r = -0.18$), but does not reach significance. Finally, log GDP per worker is essentially zero ($r = 0.04$), suggesting that national productivity adds nothing once digital intensity is accounted for. The cross-country variation in AI adoption gradients does, therefore, not seem to be a story about richer versus poorer countries in Europe; it is about digitally intensive versus digitally sparse ones, with training provision and cognitive job demands as the margins that differentiate among those with digitally ready economies.

The partial correlations should be interpreted with appropriate caution given the research design and small number of countries ($N = 35$). They identify cross-country bivariate associations with the exposure--adoption gradient conditional on digital intensity. The degrees of freedom do not support richer multivariate specifications. Albania is a consistent outlier across all panels, with a steeper slope than any country characteristic predicts; excluding it yields minimally stronger partial correlations without altering the substantive pattern.

\medskip
Taken together, the country-level analysis reinforces and extends the individual-level findings. Across 35 European countries, significant national differences in AI adoption persist even after controlling for demographics, job content, and industry structure. Where adoption is higher, it is driven specifically by workers in AI-exposed occupations, not by uniform uptake across the workforce. Beyond digital intensity, which is a fundamental prerequisite for AI software use, training provision and jobs' level of abstract tasks are the country characteristics that most strongly predict how well occupational exposure translates into adoption. The pattern is consistent across levels of analysis: at the individual level, workers with formal qualifications, in jobs with high levels of abstract task content, and with a say in organisational decisions convert exposure into uptake; at the country level, nations with widespread training systems and cognitively complex job stocks exhibit steeper exposure--adoption gradients. The implication is that generative AI adoption is not diffusing passively along occupational exposure lines, but is actively conditioned by the skill and institutional infrastructure that enables workers to act on the technological opportunities and pressures their jobs present.

\subsection{Worker-Reported Consequences of AI Adoption}
\label{sec:impacts}

The preceding sections documented who adopts generative AI and under what conditions. This extension asks whether early adoption has begun to reshape the task content of jobs. Workers in the EWCS were asked whether technology has removed some of their tasks and whether it has created new ones. At the individual level, generative AI adopters report substantially more of both: 13 percentage points ($p < 0.001$) more task displacement and 10 percentage points ($p < 0.001$) more task creation than non-adopters, conditional on overall digital tool use and a comprehensive set of fixed effects including for occupation, industry, country, and individual characteristics (see Appendix~\ref{app:micro_impacts}). However, these associations cannot distinguish between whether AI adoption causes task restructuring or whether workers in already-changing jobs are more likely to adopt. To make progress on this question, we move to an aggregate analysis that can exploit variation in AI exposure as an exogenous source.

\begin{figure}[htbp]
    \centering
    \caption{Job task restructuring and AI adoption across occupations.}
    {\includegraphics[width=0.9\textwidth]{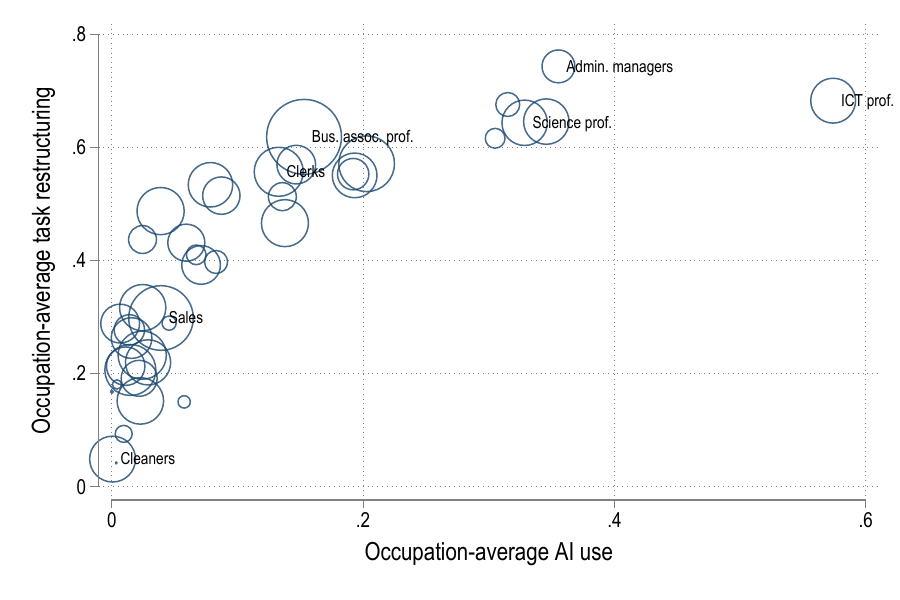} \\
    \footnotesize\textit{Notes:} Bubble size proportional to occupation
    employment. Two-digit ISCO-08 occupations. EWCS 2024.}
    \label{fig:task_restructuring}
\end{figure}

Figure~\ref{fig:task_restructuring} motivates this analysis by plotting occupation-average task restructuring against AI adoption across two-digit ISCO occupations. The association is strong (Spearman $\rho = 0.92$, $p < 0.001$): in occupations with the highest AI adoption rates, such as ICT professionals, science and engineering professionals, and administrative managers, over 60 percent of incumbents report technology-induced changes to their task content. By contrast, occupations with minimal AI use report substantially lower rates of task restructuring.

\paragraph{Empirical approach.}
To interrogate whether this association reflects a genuine effect of AI adoption on task content, the sample is restricted to workers who use a computer at work --- the at ``risk'' population  --- data are subsequently collapsed to one-digit ISCO $\times$ country cells ($N = 344$). Cell-level task displacement and creation rates are regressed on cell-level AI adoption rates, progressively adding one-digit occupation and country fixed effects. To address endogeneity in cell-level AI adoption, I construct a shift-share instrument following \textcite{Borusyak2025AInstruments}, by interacting occupation-level GAISI exposure (the ``share'') with the leave-one-out country-average AI adoption rate (the ``shift''). The instrument predicts that high-exposure occupations will exhibit disproportionately more AI adoption in countries where adoption is generally high. The approach is structurally similar to \textcite{Hampole2025ArtificialMarket}'s instrument for AI adoption, but at country rather than firm-level.  

Under the exogenous-shares interpretation of \textcite{Goldsmith-Pinkham2020BartikHow}, identification requires that occupational AI exposure does not predict task restructuring through channels other than AI adoption, conditional on the fixed effects. The design addresses several threats. The restriction to computer users ensures comparisons are drawn among workers with access to the fundamental digital infrastructure to use AI software. Occupation fixed effects absorb any tendency for high-GAISI occupations to restructure faster regardless of AI adoption. Country fixed effects absorb common macroeconomic shocks and country-level differences in digital infrastructure. Together, these rule out confounds that operate through either the occupational or country margin alone. The remaining threat is a confound that disproportionately restructures high-exposure occupations specifically in high-adoption countries, through a non-AI channel. The most plausible candidate is broader digital restructuring (e.g. to address the shift to remote working). To assess the sensitivity of the main results, we thus re-estimate the estimation model with the share of workers who work at least sometimes from home as an additional control variable. As a further check, we replace the EWCS-based shift with the change in enterprise-level generative AI deployment rates between 2021 and 2024 from the Eurostat ICT--Enterprise Survey, an independent data source, to rule out common survey measurement contamination.

\paragraph{Results.}
Table~\ref{tab:aggregate_task} reports the results. Without fixed effects (column~1), AI adoption is positively associated with both task displacement ($+0.186$, $p < 0.05$) and, more strongly, task creation ($+0.647$, $p < 0.001$). For task displacement, the coefficient collapses immediately with occupation fixed effects (column~2: $+0.034$, $n.s.$) and remains near zero with country fixed effects (column~3). For task creation, the attenuation is more gradual: the coefficient survives occupation fixed effects (column~2: $+0.401$, $p < 0.001$) but attenuates substantially with country fixed effects (column~3: $+0.205$, $p < 0.05$). The IV estimate in column~(4) cannot reject the null for either outcome, with a strong first stage ($F = 43.2$).

Two robustness checks reinforce this conclusion. Column~(5) adds the cell-level share of workers who work at least sometimes from home, a factor related to digital transformation disproportionately affecting high-skill occupations. The first stage weakens to $F = 18.7$, indicating that the instrument's power derives in part from broader digitalisation rather than AI-specific variation. The IV point estimates widen but remain null for task creation and negative for task displacement. Column~(6) replaces the EWCS-based shift with the enterprise-level change in AI deployment rates between 2021--2024 from the Eurostat ICT Enterprise Survey 2024. The first stage remains strong ($F = 32.5$), and the results are substantively identical to column~(4) for a reduced number of countries.

\begin{table}[htbp]
\centering
\begin{threeparttable}
\caption{AI adoption and task restructuring: occupation $\times$ country evidence}
\label{tab:aggregate_task}
\scriptsize
\begin{tabular*}{\textwidth}{@{\extracolsep{\fill}}l*{6}{c}}
\toprule
 & (1) & (2) & (3) & (4) & (5) & (6) \\
 & OLS & $+$Occ FE & $+$Country FE & IV & IV$+$WFH & IV (Eurostat) \\
\midrule
\multicolumn{7}{l}{\textit{Panel A: Task displacement}} \\[3pt]
AI use & $0.186^{*}$ & $0.034$ & $0.032$ & $-0.251$ & $-0.476$ & $-0.238$ \\
       & (0.069) & (0.146) & (0.089) & (0.134) & (0.275) & (0.129) \\[6pt]
\multicolumn{7}{l}{\textit{Panel B: Task creation}} \\[3pt]
AI use & $0.647^{***}$ & $0.401^{***}$ & $0.205^{*}$ & $0.100$ & $-0.006$ & $0.070$ \\
       & (0.056) & (0.101) & (0.095) & (0.164) & (0.283) & (0.225) \\[3pt]
\midrule
Occupation FE & No & Yes & Yes & Yes & Yes & Yes \\
Country FE & No & No & Yes & Yes & Yes & Yes \\
WFH share & No & No & No & No & Yes & No \\
$1^{st}$-stage F & & & & 43.2 & 18.7 & 32.5 \\
$N$ & 344 & 344 & 344 & 344 & 344 & 305 \\
\bottomrule
\end{tabular*}
\begin{tablenotes}[flushleft]
\footnotesize
\item \textit{Notes:} Unit of observation: one-digit ISCO $\times$ country cell. Sample restricted to workers using a computer at work. Cell means computed using survey weights; regressions precision weighted by cell sample size. Columns~(4) and~(5) instrument cell-level AI adoption with GAISI $\times$ leave-one-out country-average AI adoption rate. Column~(5) adds the cell-level share of workers who work at least sometimes from home. Column~(6) instruments with GAISI $\times$ the change in enterprise AI deployment rates between 2021--2024 (Eurostat ICT Enterprise Survey 2024; $N$ reduced due to country coverage). Standard errors clustered at country level.
$^{***}p<0.001$, $^{**}p<0.01$, $^{*}p<0.05$.
\end{tablenotes}
\end{threeparttable}
\end{table}

\medskip
Overall, the progressive attenuation across columns tells a consistent story. The striking descriptive association between AI adoption and task restructuring documented in Figure~\ref{fig:task_restructuring} is largely compositional: occupation fixed effects eliminate the displacement association entirely, and country fixed effects substantially weaken the task creation association. Three IV specifications, with varying ``shift'' source and controlling for remote work, confirm the null, though column~(5) suggests the instrument partly captures broader digitalisation alongside AI-specific adoption.

Two interpretations are consistent with this pattern. First, workers in already-changing jobs may select into AI adoption, generating an association that dissolves once occupational and country composition are accounted for. Second, at this early stage of diffusion, generative AI may be fitted into existing work processes and task bundles, rather than displacing or creating tasks that workers recognise as distinct. Both interpretations are consistent with the null earnings effects in Danish administrative data \parencite{Humlum2025TheWorkers,Humlum2025LaborMarket} and the modest aggregate productivity effects projected by \textcite{Acemoglu2025MacroAI}. What the analysis can say with confidence is that there is no detectable effect of early generative AI adoption on worker-perceived task composition of jobs. 

These results should be read in the context of what the identification strategy can establish. The design addresses reverse causality and absorbs occupation- and country-level confounds, but cannot fully rule out that broader digitalisation disproportionately restructures high-exposure occupations in high-adoption countries through non-AI channels. If such confounding continues to be present, it would bias the IV estimate upward, making the null more rather than less informative. What the analysis can say with confidence is that there is no detectable average marginal effect of early generative AI adoption on worker-reported technology-related task displacement or creation in this cross-section of countries. 

\section{Discussion}
\label{sec:discussion}

Generative AI is diffusing unevenly across European workplaces. Across 35 countries, adoption in 2024 ranged by an order of magnitude, from under three percent to roughly a quarter of the workforce. The central organising factor is occupational exposure: workers in occupations whose task content can be meaningfully augmented by large language models adopt at substantially higher rates in every country. But the association between exposure and adoption varies sharply across countries, and the variation is systematic. Digital intensity, the share of workers using a computer almost all of the time, is the fundamental prerequisite and the dominant raw predictor of the cross-country gradient. Beyond digital infrastructure, training provision and the abstract task intensity of the 
national job stock are the country characteristics that most strongly condition how well AI exposure translates into uptake. Log GDP per worker and the aggregate tertiary share are essentially unrelated to the gradient once digital intensity is accounted for. This refines the standard tech-diffusion story \parencite{Comin2010AnDiffusion,Keller2004InternationalDiffusion}: at very early stages of generative AI uptake across Europe, what matters is not national productivity or the human capital stock writ large, but whether the workforce is digitally integrated and supported by institutional settings that sustain absorptive capacity.

The pooled within-country analysis points in a compatible direction. Conditional on sharing the same two-digit occupation, workers who adopt are disproportionately tertiary-educated, perform more abstract tasks, and, distinctively, exercise more influence in organisational decisions. Task discretion over order, method, and pace of work correlates with adoption separately but does not survive joint estimation, suggesting that what matters is not micro-level discretion but influence over how work is organised. This finding aligns with \textcite{Bick2026MindU.S.}, who show that differences in high-performance human resource management account for a substantial share of the U.S.--Europe adoption gap. Our evidence locates the mechanism in the broader high-involvement management infrastructure that gives workers a say in how their work is structured \parencite{AppelbaumEileen2001HPWp,Felstead2010EmployeeAnalysis}.

While this paper has examined generative AI adoption at the worker level, worker use does not occur in isolation from firm deployment. The country-by-industry correlation between worker-reported adoption in the EWCS and firm-reported AI adoption in the Eurostat ICT Enterprise Survey is $r = 0.86$: across 303 industry-by-country cells, the two measures track each other closely. This convergence, together with the moderator evidence that organisational conditions shape the exposure--adoption gradient within occupations, indicates that the worker-level phenomenon we document is bound up with firm-level AI deployment, not separable from it. The evidence cannot discriminate whether firm investment enables worker use, worker use prompts firm formalisation, or both respond to common conditions such as the digital and organisational infrastructure of high-adoption industries. The evidence presented here supports the idea that the organisational features we measure --- high-involvement work practices, training systems, and digital integration --- operate as accelerators of worker AI uptake. Whether firms also exercise a gatekeeping role through explicit restrictions, as documented by \textcite{Humlum2025LaborMarket}, is a question the EWCS does not allow us to address.

A persistent gender gap cuts across these patterns. Men are significantly more likely than women in the same two-digit occupation to adopt generative AI, a gap that is concentrated in the most exposed occupations and pervasive across European countries. Importantly, the gap cannot be explained by occupational sorting or human capital endowment; it reflects a within-occupation differential in the translation of exposure into uptake. \textcite{Bick2026MindU.S.} find a narrower gap in their late-2025 U.S.-Europe data, raising the possibility that the here-observed gender gap reflects a transitional lag rather than a durable structural disparity. Whether the gap closes as the technology matures or persists through institutional mediation---the within-occupation differential and its cross-country variation (from negative in Albania to substantial in Denmark) point to institutional context rather than the technology itself---cannot be resolved with cross-sectional data. The concentration of the gap in the most exposed occupations, precisely where productivity gains from adoption are likely to materialise, makes it a first-order question for research and for employer practice.

The findings add to the emerging evidence of modest short-run aggregate effects of generative AI despite substantial adoption. \textcite{Humlum2025LaborMarket,Humlum2025TheWorkers} document rapid ChatGPT uptake in exposed Danish occupations with no detectable effects on earnings or hours, and \textcite{Acemoglu2025MacroAI} projects maximum TFP gains below one percent over a decade. The analysis of task restructuring offers a complementary null from a different empirical angle: once occupational and country composition are accounted for, there is no detectable effect of early adoption on worker-reported task displacement or creation. Adopters perceive more restructuring than non-adopters at the individual level, but the aggregate shift-share design indicates this association largely reflects the selection of workers in already-changing jobs into adoption. Both our null and the null earnings effects elsewhere are consistent with the productivity J-curve of \textcite{Brynjolfsson2021JCurves}: the organisational and individual learning required to capture productivity gains from general-purpose technologies precedes their realisation in observable outcomes. Early adoption in the 2024 EWCS likely reflects a transitional phase in which generative AI is fitted into existing task bundles and work processes, rather than actively reshaping them. Whether restructuring emerges as adoption deepens is a question only longitudinal data can answer. Such effects are likely to be institutionally mediated, as this analysis finds for adoption itself, and as earlier waves of automation suggest for economic outcomes \parencite{Lewandowski2025Auto}.

What do these findings imply for policy? The European Commission's AI strategy states that ``no one is left behind in the digital transformation'' and identifies modernisation of education and skills provision as a priority \parencite{european_commission_2018_ai_for_europe}. The presented evidence is consistent with this framing but redirects attention in one specific respect. The country-level correlate that most robustly predicts the exposure--adoption gradient, conditional on digital infrastructure, is workplace training prevalence. This is a feature of adult-learning systems, not of AI training programmes specifically: countries where workers routinely engage in employer-provided or self-financed training convert occupational exposure into adoption more efficiently. The implication is that the absorptive capacity created by established lifelong-learning infrastructure extends to generative AI, and likely will extend to derived technologies that follow it. Beyond training systems, our findings identify a need for statistical infrastructure to track AI adoption, distributional consequences, and task-level effects as they evolve. Single-wave surveys of AI use, including this one, cannot settle questions about whether the gender gap narrows, whether the exposure--adoption gradient steepens or flattens with diffusion, or whether economic consequences emerge as adoption deepens.

Several limitations qualify this study's findings. First, the EWCS adoption measure is binary, conflating occasional with intensive use. Moreover, by design, the measure picks up only AI tools salient to workers, excluding backend AI systems embedded in workflows without direct user interaction. Convergent validity with enterprise-level AI adoption across European industries and countries ($r = 0.86$) suggests that the EWCS item captures meaningful variation in workplace AI engagement at the aggregate level despite these limitations. Second, the moderation analysis identifies conditional associations, not causal effects: we cannot rule out that workers with unobserved characteristics sort into high-involvement, training-rich and abstract employment, and adopt for reasons unrelated to these features. The within-occupation identification does, however, absorb substantial selection on skills and task content and, in so doing, addresses central confounds. Third, the shift-share design on task restructuring addresses some sources of reverse causality but cannot fully rule out confounds operating through broader digitalisation of high-exposure occupations in high-adoption countries. It is likely that in the presence of such confounds, the presented estimates represent upper bounds of generative AI's effects on worker-reported task restructuring. Fourth, the 2024 EWCS captures adoption roughly eighteen months after ChatGPT's release, and the quantitative levels reported here will have shifted substantially in the interim. What is likely more durable is the conditional structure of adoption, including who adopts given exposure, and where exposure translates most efficiently into uptake, and it is this structure, rather than the levels, that constitutes the contribution of this paper. The conditional structure of adoption documented here provides the baseline against which future evidence using longitudinal (quasi-) experimental research designs will be read.

\bigskip
\printbibliography

\clearpage

\section*{Biographical note} 
Golo Henseke is an Associate Professor at UCL's Institute of Education and a co-investigator on the Skills and Employment Survey, Britain's longest-running programme tracking work quality and skills utilisation, now in its fourth decade. His applied economic research on skills, job quality, technological change, and well-being draws on large-scale survey and administrative data and has appeared in Oxford Economic Papers, ILR Review, British Journal of Industrial Relations, and New Technology, Work and Employment.

\section*{AI assistance}
LLM assistance was used for coding, drafting, and formatting in this project. After using these services, the author reviewed and edited the content as needed and takes full responsibility for the content of the publication.

\section*{Data Availability}
The 2024 European Working Conditions Survey is hosted by the UK Data Service (\hyperlink{http://doi.org/10.5255/UKDA-SN-9511-1}{SN: 9511}) and available to registered users subject to the \hyperlink{https://ukdataservice.ac.uk/app/uploads/cd137-enduserlicence.pdf}{End User Licence Agreement}.

\section*{Conflict of Interest} 
None to declare

\section*{Ethical statement}  
This research used safeguarded secondary data from the UK Data Service. The author affirms strict compliance with the End User Licence Agreement.

\clearpage
\begin{appendices}

\renewcommand{\thetable}{\thesection.\arabic{table}}
\renewcommand{\thefigure}{\thesection.\arabic{figure}}
\setcounter{table}{0}
\setcounter{figure}{0}

\section{Country Coverage}
\label{app:countries}

\setcounter{table}{0}
\setcounter{figure}{0}

The EWCS 2024 covers the following 35 countries: Austria (AT), Belgium (BE), Bulgaria (BG), Croatia (HR), Cyprus (CY), Czech Republic (CZ), Denmark (DK), Estonia (EE), Finland (FI), France (FR), Germany (DE), Greece (EL), Hungary (HU), Ireland (IE), Italy (IT), Latvia (LV), Lithuania (LT), Luxembourg (LU), Malta (MT), Montenegro (ME), Netherlands (NL), North Macedonia (MK), Norway (NO), Poland (PL), Portugal (PT), Romania (RO), Serbia (RS), Slovakia (SK), Slovenia (SI), Spain (ES), Sweden (SE), Switzerland (CH), Bosnia and Herzegovina (BA), Albania (AL), Kosovo (XK).

\section{Measurement}
\setcounter{table}{0}
\setcounter{figure}{0}

\subsection{Convergent Validity of Worker-Reported AI Use}
\label{app:ss:convergent_validity}
\begin{figure}[htbp]
\centering
    \caption{Convergent validity: worker-reported and enterprise-reported AI use.}
    \label{app:fig:validation}
    \includegraphics[width=\textwidth]{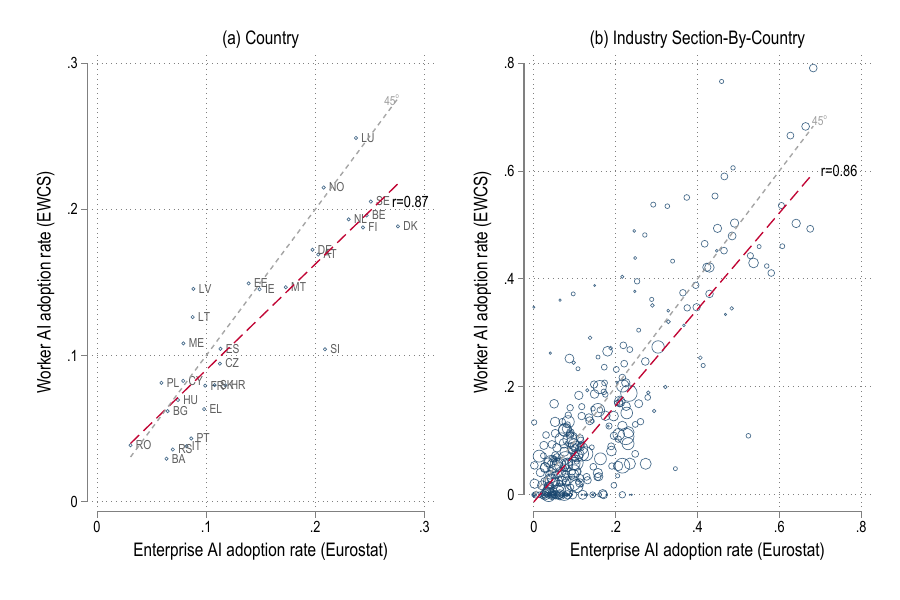}
    \begin{minipage}{\textwidth}
    \vspace{0.3em}
    \footnotesize
    \textit{Notes:} In (a) each point represents a country cell ($N_{c} = 31$), and in (b) an industry (NACE Level~1)-by-country cell ($N_{ic} = 303$). The horizontal axis plots cells by the share of enterprises using at least one AI technology (EU-ICT-Firm 2024; doi:~\href{https://doi.org/10.2908/ISOC_EB_AIN2}{10.2908/ISOC\_EB\_AIN2}). The vertical axis plots cells by the weighted share of workers who report using generative AI at work (EWCS 2024). Cells are precision-weighted by the number of EWCS respondents. The fitted line is from precision-weighted OLS on firm AI adoption. Pearson $r = 0.87$ (a) and $r = 0.86$ (b), respectively (both $p < .001$).
\end{minipage}
\end{figure}

Figure~\ref{app:fig:validation} compares worker-reported generative AI adoption from the EWCS with enterprise-reported AI use from the EU-ICT-Firm Survey, both measured in 2024. The two measures differ in relevant aspects: the Eurostat survey asks firms whether they use any of seven AI technologies (perform analysis of written language, convert spoken language into machine-readable format, generate written language, spoken language or code, identify objects or persons based on images, machine learning, to automate workflow or assist decision-making, enable physical movement of machines), covering a broader range of applications than the EWCS question, which anchors generative tools through named examples (ChatGPT, Midjourney, etc.). Respondents also differ (ICT managers or high-level ICT professionals in the Eurostat survey versus individual workers in the EWCS), as does the mode of data collection (online business survey versus face-to-face interview). Moreover, EU-ICT-Firm is limited to organisations with 10 or more employees, excluding about a quarter of employees in smaller workplaces. Despite these differences, the Pearson correlation across $N = 31$ countries is $r = 0.87$ (plot (a)) and $r = 0.86$  across  $N = 303$ industry-by-country cells (plot (b)), indicating strong convergent validity. In plot (a), most countries cluster around the fitted line, but notable deviations exist. Slovenia stands out with an enterprise adoption rate comparable to Germany and Austria, yet worker-reported use is roughly half as high. A broader group of Bosnia and Herzegovina, Italy, Portugal, and Serbia also falls below the fitted line, suggesting a broader pattern in which enterprise-reported adoption in these economies is not as strongly correlated with worker-level engagement with AI tools as in other European countries. In plot (b), cells where both measures report low adoption dominate the distribution, consistent with the early stage of AI diffusion across most sectors and countries. As in (a), some cells show high enterprise-reported but low worker-reported AI use, likely reflecting backend AI systems that are invisible to individual workers. Conversely, a smaller number of cells show high worker-reported use but modest enterprise-reported adoption, consistent with informal, employee-initiated use of generative AI tools that may not register in firm-level surveys.

\subsection{Factor Analysis Loadings}
\label{app:factor}

Table~\ref{tab:factor} reports the rotated factor loadings from the principal components analysis used to construct the Abstract and Routine task content variables.

\begin{table}[H]
\centering
\begin{threeparttable}
\caption{Rotated factor loadings: Abstract and Routine Tasks}
\label{tab:factor}
\scriptsize
\begin{tabular*}{\textwidth}{@{\extracolsep{\fill}}l*{3}{c}}
\toprule
Item & Factor 1: Abstract  & Factor 2: Routine & Uniqueness \\
\midrule
Encounters unforeseen problems & 0.610 & 0.048 & 0.626 \\
Performs complex tasks         & 0.751 & 0.050 & 0.433 \\
Learning new things            & 0.726 & $-$0.007 & 0.472 \\
Uses foreign languages         & 0.495 & 0.001 & 0.755 \\
Difficult decisions            & 0.729 & $-$0.028 & 0.468 \\
Monotonous tasks               & 0.026 & 0.563 & 0.682 \\
Repetitive tasks ($<$1 min)   & $-$0.008 & 0.821 & 0.326 \\
Repetitive tasks ($<$10 min)  & 0.024 & 0.836 & 0.300 \\
\midrule
Eigenvalue (post-rotation)     & 2.241 & 1.696 & \\
Variance explained             & 28.0\% & 21.2\% & \\
Cumulative                     & 28.0\% & 49.2\% & \\
\bottomrule
\end{tabular*}

\begin{tablenotes}[flushleft]
\footnotesize
\item \textit{Notes:} Principal components factor analysis with varimax rotation. All items recoded to binary (1=yes, 0=no) prior to analysis. EWCS 2024.
\end{tablenotes}
\end{threeparttable}
\end{table}

\section{Robustness: Alternative Exposure Measure}
\label{app:emmr}
\setcounter{table}{0}
\setcounter{figure}{0}

Table~\ref{tab:emmr_robust} replicates the pooled exposure-adoption models from Table~\ref{tab:pooled} using the human-rater $\beta$--exposure score from \textcite{Eloundou2024GPTs} in place of our derived measure. The Eloundou et al.\ measure is constructed from expert assessments of O*NET tasks and mapped to two-digit ISCO occupations; it is expressed as a proportion of exposed tasks (mean $= 0.293$, SD $= 0.157$) rather than standardised. The pattern of results is substantively identical: occupational exposure strongly predicts adoption in the baseline specification and attenuates by roughly half with progressive controls, confirming that the exposure-adoption gradient is not an artefact of the specific index construction. To compare magnitudes, the implied effect of a one-standard-deviation increase in Eloundou et al.\ exposure is $0.157 \times 0.595 = 9.3$~pp in column~(1) and $0.157 \times 0.258 = 4.1$~pp in column~(3), closely comparable to the GAISI estimates of 10.1~pp and 4.9~pp, respectively.

\begin{table}[htbp]
\centering
\begin{threeparttable}
\caption{Occupational exposure and generative AI adoption: Eloundou et al.\ human-rater scores}
\label{tab:emmr_robust}
\small
\begin{tabular*}{\textwidth}{@{\extracolsep{\fill}}l*{3}{c}}
\toprule
 & (1) & (2) & (3) \\
 & Demographics & $+$Skills & $+$Organisation \\
\midrule
\multicolumn{4}{l}{\textit{DV: AI use at work (0/1)}} \\[3pt]
Eloundou et al.\ exposure & 0.595$^{***}$ & 0.335$^{***}$ & 0.258$^{***}$ \\
 & (0.048) & (0.047) & (0.045) \\[3pt]
\midrule
Country FE          & Yes & Yes & Yes \\
Foreign-born, Age$\times$Sex   & Yes & Yes & Yes \\
Education, seniority & No & Yes & Yes \\
Computer use         & No & Yes & Yes \\
Employment status   & No & No & Yes \\
Workplace size     & No & No & Yes \\
Industry FE         & No  & No  & Yes \\
Observations        & 35{,}130 & 35{,}130 & 35{,}130 \\
$R^2$               & 0.128 & 0.169 & 0.225 \\
\bottomrule
\end{tabular*}
\begin{tablenotes}
\footnotesize
\item \textit{Notes:} Linear probability model estimates replicating Table~\ref{tab:pooled} with the \textcite{Eloundou2024GPTs} overall human-rater exposure score (proportion scale, mean $= 0.293$, SD $= 0.157$) in place of GAISI. Controls as in Table~\ref{tab:pooled}. Survey weights applied (\texttt{calweight}). Standard errors clustered at country$\times$two-digit occupation. $^{***}p<0.001$, $^{**}p<0.01$, $^{*}p<0.05$.
\end{tablenotes}
\end{threeparttable}
\end{table}

\section{Supplementary Analyses}
\setcounter{table}{0}
\setcounter{figure}{0}

\subsection{Exposure--Adoption Gradient by Country}

\begin{figure}[htbp]
    \centering
    \caption{Country-specific GAISI slopes with 95\% confidence intervals.\\
    \includegraphics[width=0.9\textwidth]{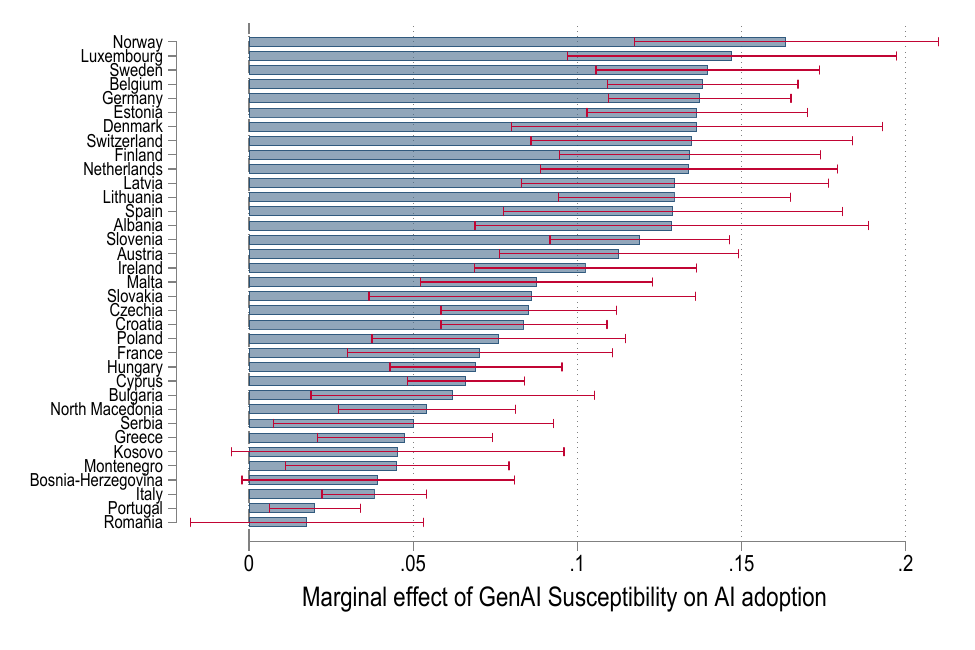} \\
    \footnotesize\textit{Notes:} Marginal effects of the GenAI Susceptibility Index on AI adoption probability, estimated separately by country from a fully interacted LPM. Bars are precision-weighted point estimates; caps indicate 95\% confidence intervals.}
    \label{app:fig:gaisi_slopes}
\end{figure}

Figure~\ref{app:fig:gaisi_slopes} reports the marginal effect of occupational exposure (GAISI) on the probability of generative AI adoption, estimated separately for each of the 35 EWCS countries from a fully interacted linear probability model (Equation~\ref{eq:pooled} interacted with country indicators). Bars show point estimates; caps show 95\% confidence intervals. Slopes are positive in every country and significant at conventional levels in 32 of 35; the three exceptions --- Romania, Bosnia and Herzegovina, and Kosovo --- have low overall adoption rates and correspondingly imprecise slope estimates rather than systematically flat gradients. The magnitude of the slope varies by roughly a factor of nine across the sample, from around 0.02 in Romania and Portugal to 0.16 in Norway.

\subsection{Individual-Level AI Adoption and Task Impacts}
\label{app:micro_impacts}

Table~\ref{tab:micro_impacts} reports individual-level associations between generative AI adoption and self-reported task displacement and creation, restricting the sample to workers who use a computer at work. 

\begin{table}[H]
\centering
\begin{threeparttable}
\caption{Generative AI adoption and self-reported task impacts (individual level)}
\label{tab:micro_impacts}
\small
\begin{tabular*}{0.9\textwidth}{@{\extracolsep{\fill}}lcc}
\toprule
 & (1) & (2) \\
 & Tasks Displaced & Tasks Created \\
\midrule
AI use & $0.129^{***}$ & $0.101^{***}$ \\
       & (0.021) & (0.021) \\[3pt]
\midrule
Observations & 26{,}104 & 26{,}104 \\
\bottomrule
\end{tabular*}
\begin{tablenotes}[flushleft]
\footnotesize
\item \textit{Notes:} Linear probability model estimates. Sample restricted to workers using a computer at work. Controls: frequency of computer use, wearable device use, and a summative index of new digital tool use at work (cobots, electronic workspaces or cooperation platforms, online meeting tools). Fixed effects: country, age$\times$sex, foreign-born, educational attainment, two-digit occupation, industry division, workplace size, employment status, and part-time. Survey weights (\texttt{calweight}). Standard errors clustered at country$\times$two-digit occupation. $^{***}p<0.001$, $^{**}p<0.01$, $^{*}p<0.05$.
\end{tablenotes}
\end{threeparttable}
\end{table}

\end{appendices}

\end{document}